\documentclass[a4paper,11pt]{article}
\usepackage{pos}

\newcommand{\be}{\begin{equation}}
\newcommand{\ee}{\end{equation}}
\newcommand{\bea}{\begin{eqnarray}}
\newcommand{\eea}{\end{eqnarray}}
\newcommand{\ei}{\end{itemize}}
\newcommand{\ben}{\begin{enumerate}}
\newcommand{\een}{\end{enumerate}}
\newcommand{\bt}{\begin{tabbing}}
\newcommand{\et}{\end{tabbing}}
\newcommand{\nn}{\nonumber}

\newcommand{\dt}{{\Delta t}}
\newcommand{\mbphys}{{m_{b, \rm phys}}}
\newcommand{\Mpiphys}{{M_{\pi, \rm phys}}}

\title{
   \begin{picture}(0,0)(0,0)%
   \put(350,75){\makebox(0,0)[l]{\textnormal{\normalsize KEK-CP-392}}}%
   \end{picture}
   Heavy flavor physics from lattice QCD
}
\ShortTitle{Heavy flavor physics from lattice QCD}

\author*[a,b,c]{Takashi Kaneko}

\affiliation[a]{
  Theory Center, Institute of Particle and Nuclear Studies,
  High Energy Accelerator Research Organization(KEK), Ibaraki 305-0801, Japan
}
\affiliation[b]{
  School of High Energy Accelerator Science,
  The Graduate University for Advanced Studies (SOKENDAI), Ibaraki 305-0801,
  Japan
}
\affiliation[c]{
  Kobayashi-Maskawa Institute for the Origin of Particles and the Universe,
  Nagoya University, Aichi 464-8602, Japan  
}

\emailAdd{takashi.kaneko@kek.jp}

\abstract{
  We review recent progress on heavy flavor physics from lattice QCD.
}

\FullConference{%
  The 39th International Symposium on Lattice Field Theory (Lattice2022),\\
  8-13 August, 2022 \\
  Bonn, Germany 
}

\begin{document}
\maketitle



\section{Introduction}


Heavy flavor physics provides an important testing ground of
the Standard Model (SM) through various flavor changing processes
of heavy hadrons.
Indeed, theoretical and experimental investigations of $B$ meson decays
have reported intriguing tensions between the SM and experiments,
so-called $B$ anomalies,
as hints of new physics beyond the SM.
These include tensions on the decay rate ratio
$R(D^{(*)})\!=\!\Gamma(B\!\to\!D^{(*)}\tau\nu)/\Gamma(B\!\to\!D^{(*)}\ell\nu)$
($\ell\!=\!e,\mu$) 
describing the lepton flavor universality violation (LFUV),
the differential decay rate of the $B\!\to\!K\mu\mu$ decay
and the angular distribution of the $B\!\to\!K^*\mu\mu$ decay.


These anomalies could be established as evidence of new physics
together with on-going experiments,
namely LHCb at CERN and SuperKEKB/Belle II at KEK.
They play complementary roles in the search for new physics:
LHCb accumulates large samples both for $B$ and $B_s$ mesons
generated through high-energy proton collisions,
whereas 
Belle II is an $e^+ e^-$ collider experiment
of high efficiency and purity 
and, hence, advantageous in $B$ decays
to final states with neutrinos and/or multiple photons.
Their physics run started in 2019
to accumulate fifty times more data
than the previous KEKB/Belle experiment by the early 2030's.
On the other hand, 
LHCb recently started Run 3,
and a high luminosity upgrade (HL-LHC) is also planned.


\begin{figure}[tb]
  \centering
  \includegraphics[angle=0,width=0.6\linewidth,clip]%
                  {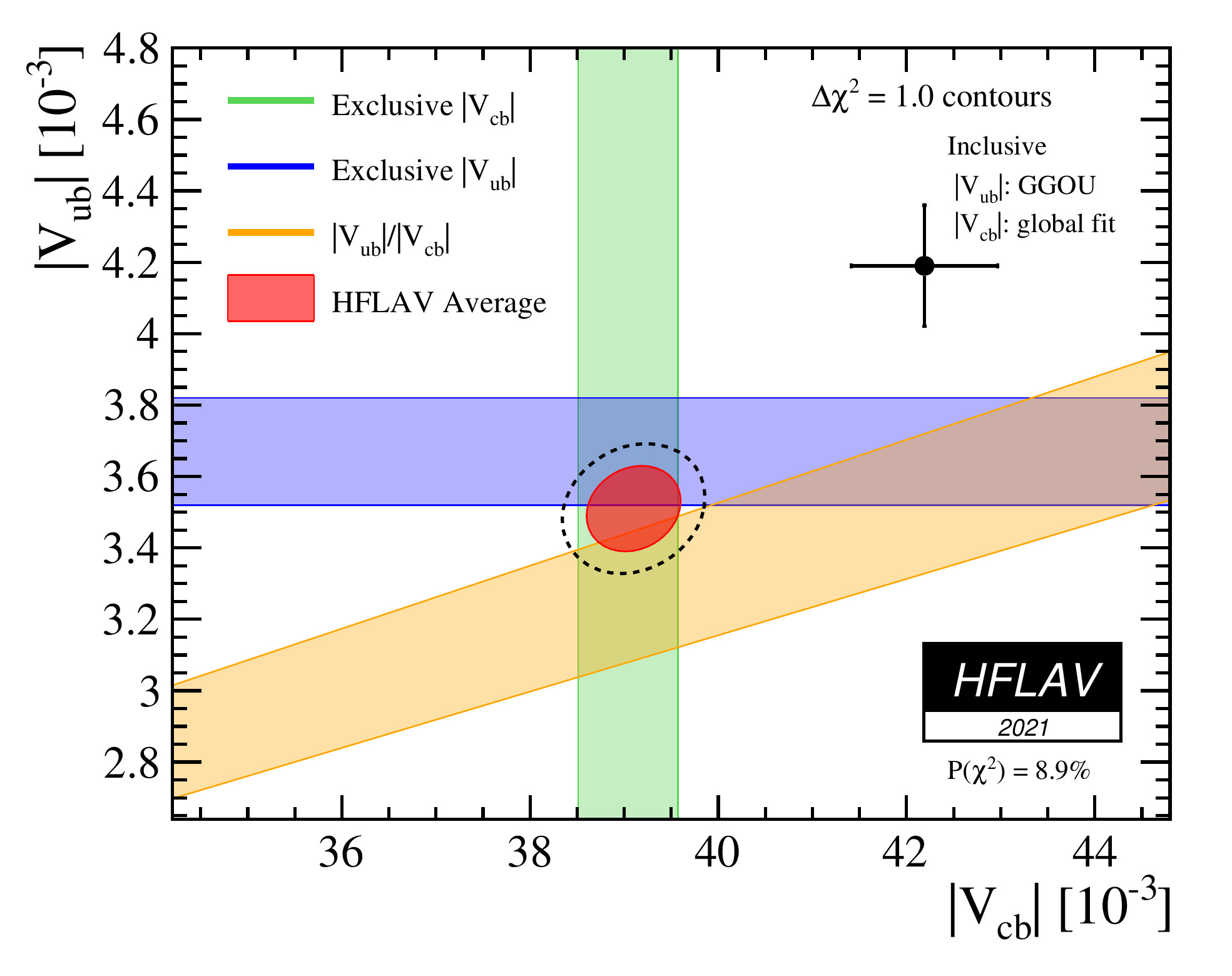}
  \vspace{-5mm}                  
  \caption{
    Recent estimate of CKM elements from exclusive (bands)
    and inclusive (black circle) decays (figure from Ref.~\cite{HFLAV22}).
    The blue horizontal band shows
    $|V_{ub}|$ determined from $B\!\to\!\pi\ell\nu$,
    whereas the green vertical band is $|V_{cb}|$ determined 
    from the $B\!\to\!D^{(*)}\ell\nu$ and $B_s\!\to\!D_s^{(*)}\ell\nu$ decays.
    The LHCb estimate of $|V_{ub}|/|V_{cb}|$ from
    $\Lambda_b\!\to\!p\mu\nu$ and $B_s\!\to\!K\mu\nu$ is also plotted
    by the slanted band. Their average (red region) is to be compared
    with the black circle from the inclusive decays.
  }
  \vspace{0mm}
  \label{fig:intro:vub_vcb}
\end{figure}

However, there has been a long-standing problem in the determination of
the Cabibbo-Kobayashi-Maskawa (CKM) matrix elements $|V_{ub}|$ and $|V_{cb}|$.
As shown in Fig.~\ref{fig:intro:vub_vcb},
there is a tantalizing $\approx\!3\,\sigma$ tension between
analyses of the exclusive decays with the specified final state hadron(s)
and the inclusive decays without such specification.
While the tension can be explained by introducing
a higher-dimensional tensor-type four fermion interaction beyond the SM,
it largely distorts the $Z\!\to\!b\bar{b}$ decay rate,
which has been precisely measured~\cite{RD:NP:CP}.
Therefore, it is likely that
the tension is not due to new physics,
but theoretical and/or experimental uncertainty has not been fully understood.
The largest theoretical uncertainty generally comes from
relevant hadronic matrix elements. 


Lattice QCD is a powerful framework to non-perturbatively calculate
the hadronic matrix elements.
There has been steady progress in the precise calculation of 
the so-called ``gold-plated quantities'',
namely matrix elements for decays to the final state
with at most one hadron, which is stable in QCD.
In addition, the past several years have witnessed
challenges to the non-gold-plated processes
especially for the inclusive decays relevant
to the $|V_{ub}|$ and $|V_{cb}|$ tensions.
In this article, we review such recent progress on the $B$ and $D$ meson decays.
We refer the readers to Ref.~\cite{FLAG5}
by Flavour Lattice Averaging Group (FLAG) 
for a comprehensive review and their world-average
mainly on the gold-plated quantities.
We also leave detailed discussions on the $B\!\to\!D^{(*)}\ell\nu$ decays
to a dedicated review talk by Vaquero
at this conference~\cite{B2Dstar:lat22}.


\section{Exclusive semileptonic decays}

\subsection{$B\!\to\!\pi\ell\nu$ decay}


The $B\!\to\!\pi\ell\nu$ decay involving the light leptons $\ell\!=\!e, \mu$
provides the conventional determination of $|V_{ub}|$.
As mentioned in the introduction, however,
this shows $\gtrsim 2~\sigma$ (12\,\%) tension
with the inclusive analysis. 
This is a CKM suppressed process with the branching fraction of 
${\mathcal B}(B\!\to\!\pi\ell\nu)\!\sim\!1.5 \times 10^{-4}$
to be compared with ${\mathcal B}(B\!\to\!D^{(*)}\ell\nu)$ of 2\,--\,5\,\%.
The accuracy of the previous Belle measurement is, therefore, limited
by the statistics.
It is expected to be largely improved by Belle II
with the aimed integrated luminosity of $\int L \!\sim\! 50\,ab^{-1}$,
which is fifty times larger than Belle.
Actually, the ``Belle II Physics Book''
by Belle II-Theory Interface Platform (B2TiP)
reported that the accuracy of $|V_{ub}|$ would be limited
by the uncertainty of lattice QCD at an early stage of Belle II
around at $\int L \!\approx\!10\,ab^{-1}$.
%
%
In addition, $B\!\to\!\pi\tau\nu$ may provide a hint of new physics
through the LFUV ratio 
$R(\pi)\!=\!\Gamma(B\!\to\!\pi\tau\nu)/\Gamma(B\!\to\!\pi\ell\nu)$
($\ell\!=\!e,\mu$),
which is expected to be measured by Belle II
with an accuracy of $\approx\!14$\,\%.
It is, therefore, an urgent task 
to improve the lattice calculation of the $B\to\pi\ell\nu$ form factors.
A target accuracy is a few \,\% or better in five years.


\begin{figure}[t]
  \centering
  \begin{minipage}[h]{0.454\linewidth}                 
    \includegraphics[angle=0,width=1.0\linewidth,clip]%
                    {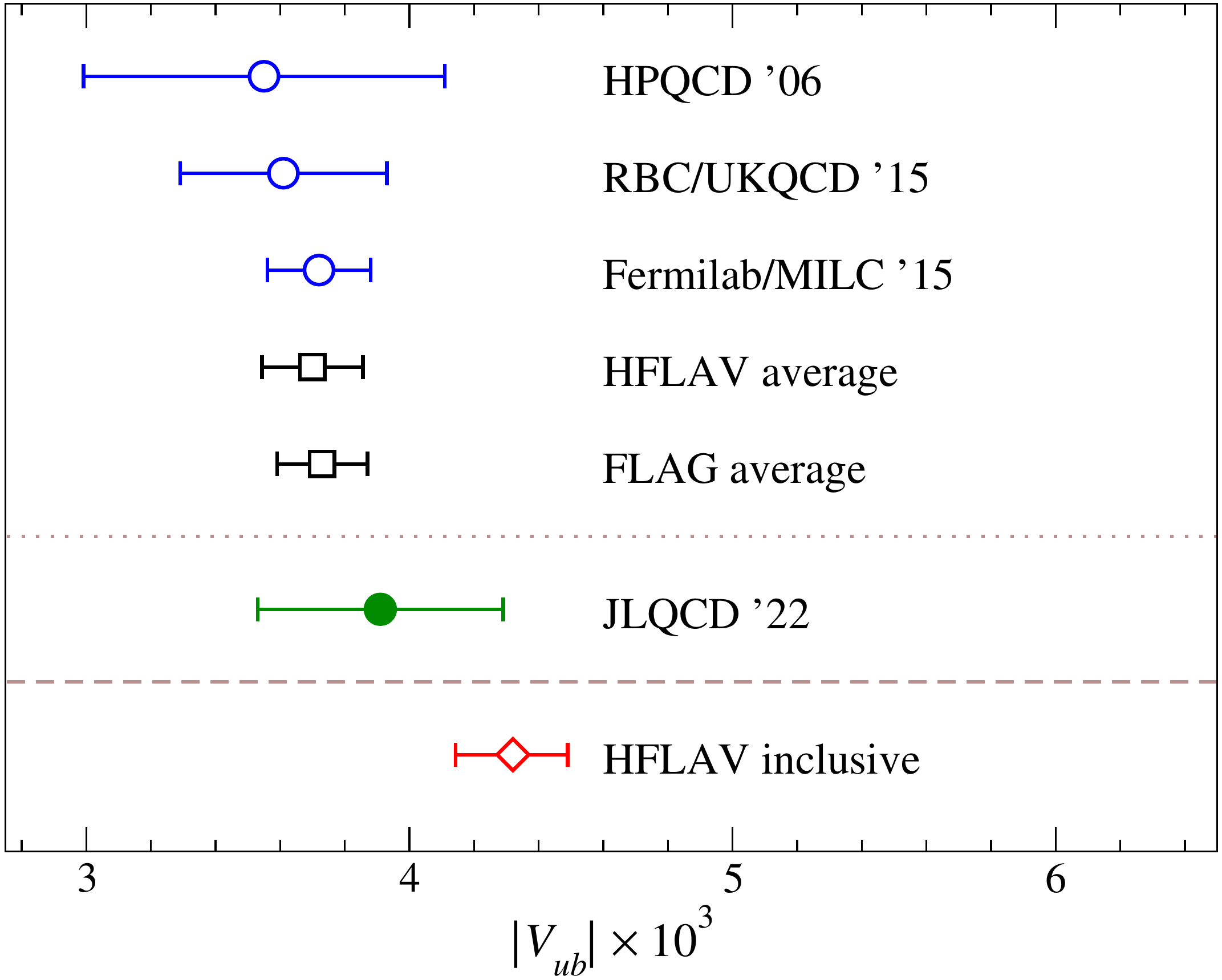}
  \end{minipage}
  \hspace{1.5mm}
  \begin{minipage}[h]{0.524\linewidth}
    \includegraphics[angle=0,width=1.0\linewidth,clip]%
                    {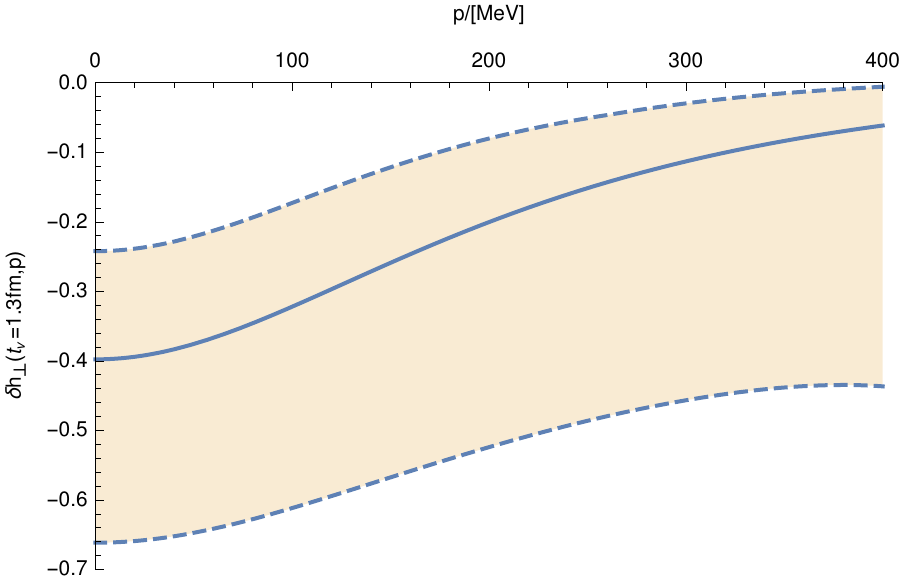}
  \end{minipage}
  \vspace{-2mm}                  
  \caption{
    Left: comparison of recent estimate of $|V_{ub}|$ from $B\!\to\!\pi\ell\nu$.
    Blue open circles show previous estimates
    by the HPQCD~\cite{B2pi:Nf3:HPQCD}, RBC/UKQCD~\cite{B2pi:Nf3:RBC/UKQCD}
    and Fermilab/MILC~\cite{B2pi:Nf3:Fermilab/MILC} Collaborations,
    whereas black open squares are obtained by
    HFLAV~\cite{HFLAV22} and FLAG~\cite{FLAG5} using these lattice data.
    They are compared with recent JLQCD's estimate~\cite{B2pi:Nf3:JLQCD}
    (filled green circle),
    and $|V_{ub}|$ from the inclusive decay (red open diamond)~\cite{HFLAV22}.
    Right: relative shift of $B\!\to\pi$ form factor $h_\perp$
    as a function of pion momentum $p$
    (figure from Ref~\cite{B*pi:ChPT:Broll:Lat22}).
    The propagating time of the intermediate $B^*\pi$ state is fixed to 1.3~fm.
  }
  \vspace{-3mm}
  \label{fig:SlD:Vub}
\end{figure}

The recent status of the determination of $|V_{ub}|$ is summarized
in the left panel of Fig.~\ref{fig:SlD:Vub}.
Previous studies by the HPQCD~\cite{B2pi:Nf3:HPQCD},
RBC/UKQCD~\cite{B2pi:Nf3:RBC/UKQCD} and
Fermilab/MILC~\cite{B2pi:Nf3:Fermilab/MILC} Collaborations
employed heavy quark actions based on effective field theories,
such as the heavy quark effective theory (HQET)
or non-relativistic QCD,
to simulate around the physical bottom quark mass $\mbphys$.
The best precision of $\approx\!4$\,\% has been achieved by Fermilab/MILC
by simulating the lattice cutoffs $\lesssim\!4.4$~GeV,
pion masses down to $M_\pi\!\approx\!180$\,MeV
and by using the so-called Fermilab approach,
namely a HQET re-interpretation of the Wilson quark action~\cite{FermilabApp}.
Recently, the JLQCD Collaboration calculated
the form factors
through the relativistic approach
using the M\"obius domain-wall action~\cite{MoebiusDWF}
for all relevant quark flavors~\cite{B2pi:Nf3:JLQCD}.
Their accuracy is typically 10\,\%
at similar lattice cutoffs
but with larger $M_\pi\!\gtrsim\!230$\,MeV.
As seen in Fig.~\ref{fig:SlD:Vub}, therefore,
the recent world averages of $|V_{ub}|$ have been dominated
by the Fermilab/MILC study about seven years ago.


The uncertainty of the Fermilab/MILC, RBC/UKQCD and JLQCD calculations
mainly comes from the statistics and continuum-chiral extrapolation.
Controlling the chiral extrapolation is not easy for $B\!\to\!\pi\ell\nu$,
because i) the chiral logarithm involves
the $B^* B\pi$ coupling~\cite{B2pi:ChPT:BPZ,B2pi:ChPT:BJ},
which can not be fixed from chiral symmetry,
and ii), in contrast to $B\!\to\!D^{(*)}\ell\nu$,
the logarithm is not suppressed
by heavy quark symmetry~\cite{B2Dstar:ChPT:RW,B2Dstar:ChPT:S}.
Therefore, we need high statistics simulations
close to the physical pion mass $\Mpiphys$
to achieve the target accuracy comparable to the Belle II measurement.
Fermilab/MILC is pursuing a relativistic calculation
using the highly improved staggered quark (HISQ) action
on the MILC gauge ensembles covering the physical point $\Mpiphys$
as well as the lattice cutoff $\lesssim\!6.6$~GeV,
where they can directly simulate
the physical bottom mass $\mbphys$~\cite{B2pi:Nf3:Fermilab/MILC:lat22}.
We also note that the RBC/UKQCD~\cite{B2pi:Nf3:RBC/UKQCD:lat21}
and JLQCD studies are being updated.


However, there has been a concern about the (near-)physical point calculations
of $B$ meson observables:
multi-particle states with additional pions give rise to 
non-negligible contamination as the pion mass decreases. 
Such excited state contamination to nucleon form factors
has been carefully studied in Ref.~\cite{Npi},
and similar contamination to $B$ meson observables
has been previously suggested by Hashimoto
at Lattice 2018~\cite{HeavyFlv:Hashimoto:Lat18}.
At this conference,
B\"ar~\cite{B*pi:ChPT:Baer:Lat22} and Broll~\cite{B*pi:ChPT:Broll:Lat22}
reported on the $B^*\pi$ state contamination to $B$ meson observables
within heavy meson chiral perturbation theory (ChPT)
in the static limit.
At the next-to-leading order (NLO),
the relevant low energy constants (LECs) are
those characterizing the $B$ meson interpolating field and
heavy-light vector current denoted by $\beta_1$ and $\beta_2$, respectively,
as well as a linear combination of LECs in the NLO Lagrangian $\gamma$.
Since these LECs are with the mass dimension $-1$
and not known,
a conservative bound from a naive dimensional analysis is assumed as 
$-\Lambda_\chi^{-1} \! \leq \! \beta_{1,2}, \gamma \! \leq \! \Lambda_\chi^{-1}$,
where $\Lambda_\chi\!\sim\!1$\,GeV is the cutoff scale of ChPT.
The right panel of Fig.~\ref{fig:SlD:Vub} shows
the relative shift $\delta h_\perp$ to the form factor 
$h_\perp\!=\!\langle \pi(p)|V_k|B(0)\rangle/\sqrt{2M_B} p_k$
in the HQET convention.
They found volume-enhanced diagrams leading to
a sizable contamination $\delta h_\perp$.
Note that the differential decay rate for
the light lepton channels $B\!\to\!\pi e \nu, \pi\mu\nu$
is described by the vector form factor $f_+$,
a large fraction of which comes from $h_\perp$.
Therefore,
a more detailed investigation is necessary
towards a precision determination of $|V_{ub}|$.
To this end,
the authors also proposed how to determine the NLO LECs
from three- and two-point functions on the lattice.


\begin{figure}[t]
  \centering
  \begin{minipage}[h]{0.487\linewidth}                 
    \includegraphics[angle=0,width=1.0\linewidth,clip]%
                    {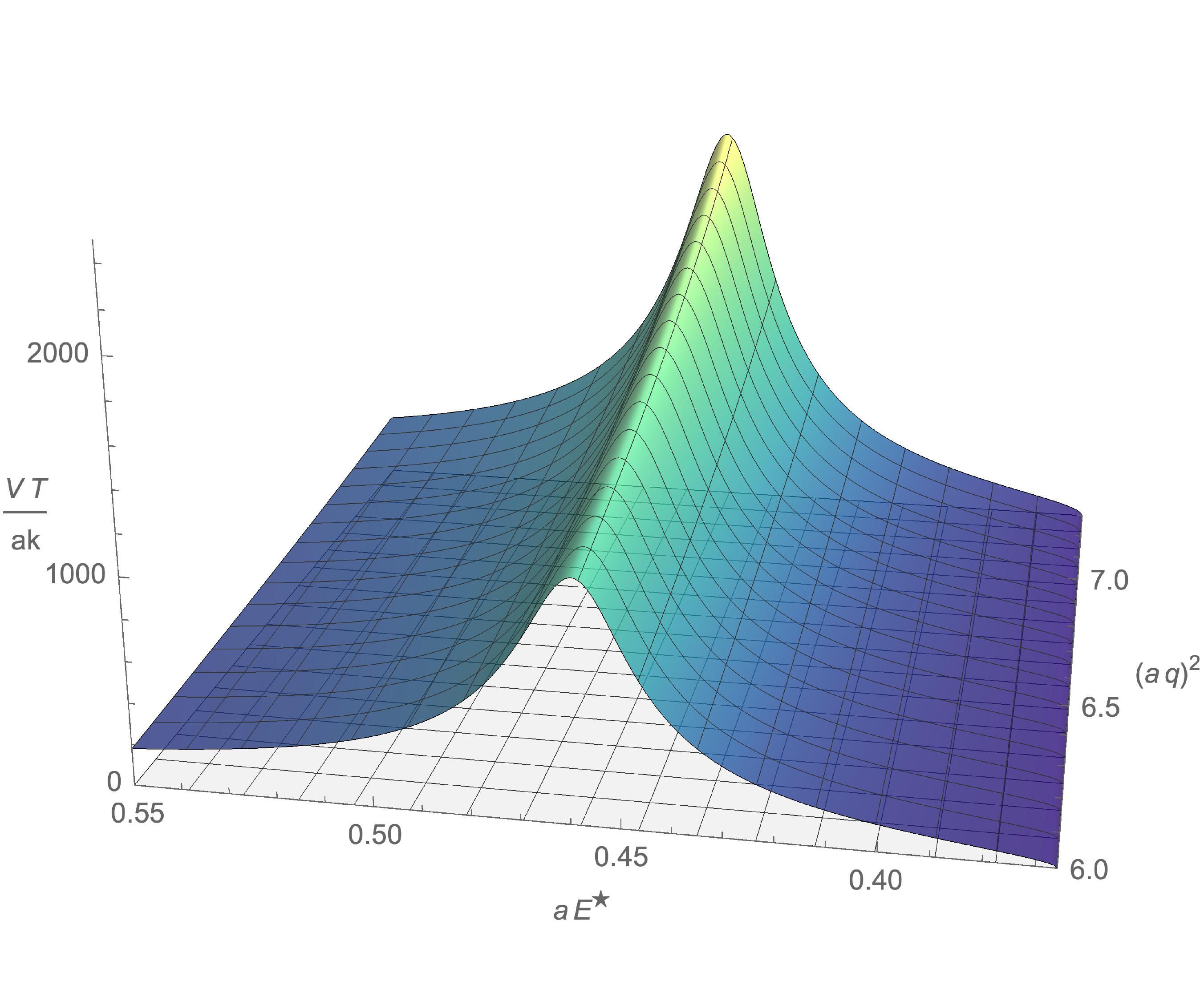}
  \end{minipage}
  \begin{minipage}[h]{0.507\linewidth}
    \includegraphics[angle=0,width=1.0\linewidth,clip]%
                    {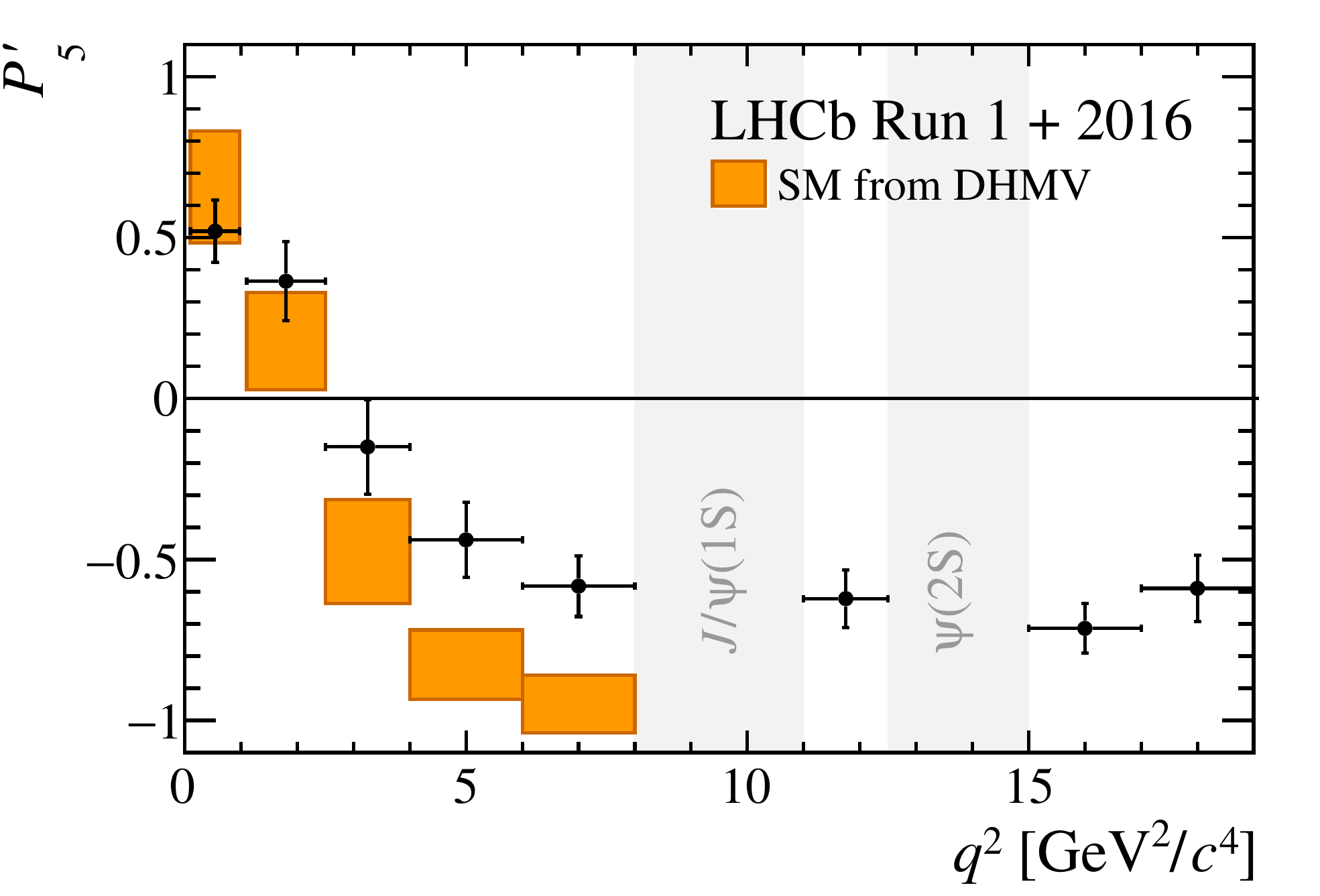}
  \end{minipage}
  \vspace{-2mm}                  
  \caption{
    Left: $B\!\to\!\rho(\to\pi\pi)\ell\nu$ transition amplitude ${\mathcal A}$
    through weak vector current (figure from Ref.~\cite{B2rho:Nf3:LMPNPPR}).
    The momentum transfer is denoted by $q^2$, 
    and $E^\star$ represents the invariant mass of the $\pi\pi$ state.
    Right: recent LHCb result for angular observable $P_5^\prime$
    for $B\!\to\!K^*\ell\ell$ as a function of momentum transfer $q^2$
    (figure from Ref.~\cite{B2Kll:LHCb}).
    The black symbols show the LHCb result, which are compared with
    the orange band
    representing a SM prediction from Ref.~\cite{B2Kll:P5p:DHMV} 
  }
  \vspace{-3mm}
  \label{fig:SlD:B2rho}
\end{figure}

The $B\!\to\!\rho\ell\nu$ decay is expected to provide
an independent determination of $|V_{ub}|$
to resolve the tension with the inclusive decay.
It also serves as new physics probe complementary to $B\!\to\!\pi\ell\nu$,
since interactions with odd intrinsic parity can contribute.
Leskovec reported an interesting progress
on this non-gold-plated mode~\cite{B2rho:Nf3:LMPNPPR}.
They simulate $N_f\!=\!2+1$ QCD
with the clover light quarks and Fermilab $b$ quarks
on a (3.6\,fm)$^3$ box at $M_\pi\!\approx\!320$~MeV,
where $\rho$ can decay into $\pi\pi$.
In contrast to the previous quenched study~\cite{B2rho:Nf0:UKQCD},
they calculate $B$ meson three-point functions
with sink operators of the single $B$ state as well as $I\!=\!1$ $\pi\pi$ state
to extract the ground-state contribution by generalized eigenvalue problem.
Then, finite volume matrix elements of the weak vector and axial currents
are connected to the infinite volume transition amplitudes
based on the formalism in Refs.~\cite{FV:1to2:BHW,FV:1to2:BDL}.
The left panel of Fig~\ref{fig:SlD:B2rho} shows their estimate of
the $B\!\to\!\rho(\to\pi\pi)\ell\nu$ transition amplitude $\mathcal A$
through the weak vector current.
While only the central value is available from
the simulation at single combination of $a$ and $M_\pi$,
this is an encouraging progress towards an independent determination
of $|V_{ub}|$ from $B\!\to\!\rho\ell\nu$.
It is also interesting to extend this approach to the $B\!\to\!K^*\ell\ell$ decay,
for which more than 3\,$\sigma$ tension in its angular distribution
persists for about ten years as shown in the right panel of
Fig.~\ref{fig:SlD:B2rho}.

\subsection{$B_s$ and $B_c$ meson decays}

\begin{figure}[t]
  \centering
  \begin{minipage}[h]{0.467\linewidth}                 
    \includegraphics[angle=0,width=1.0\linewidth,clip]%
                    {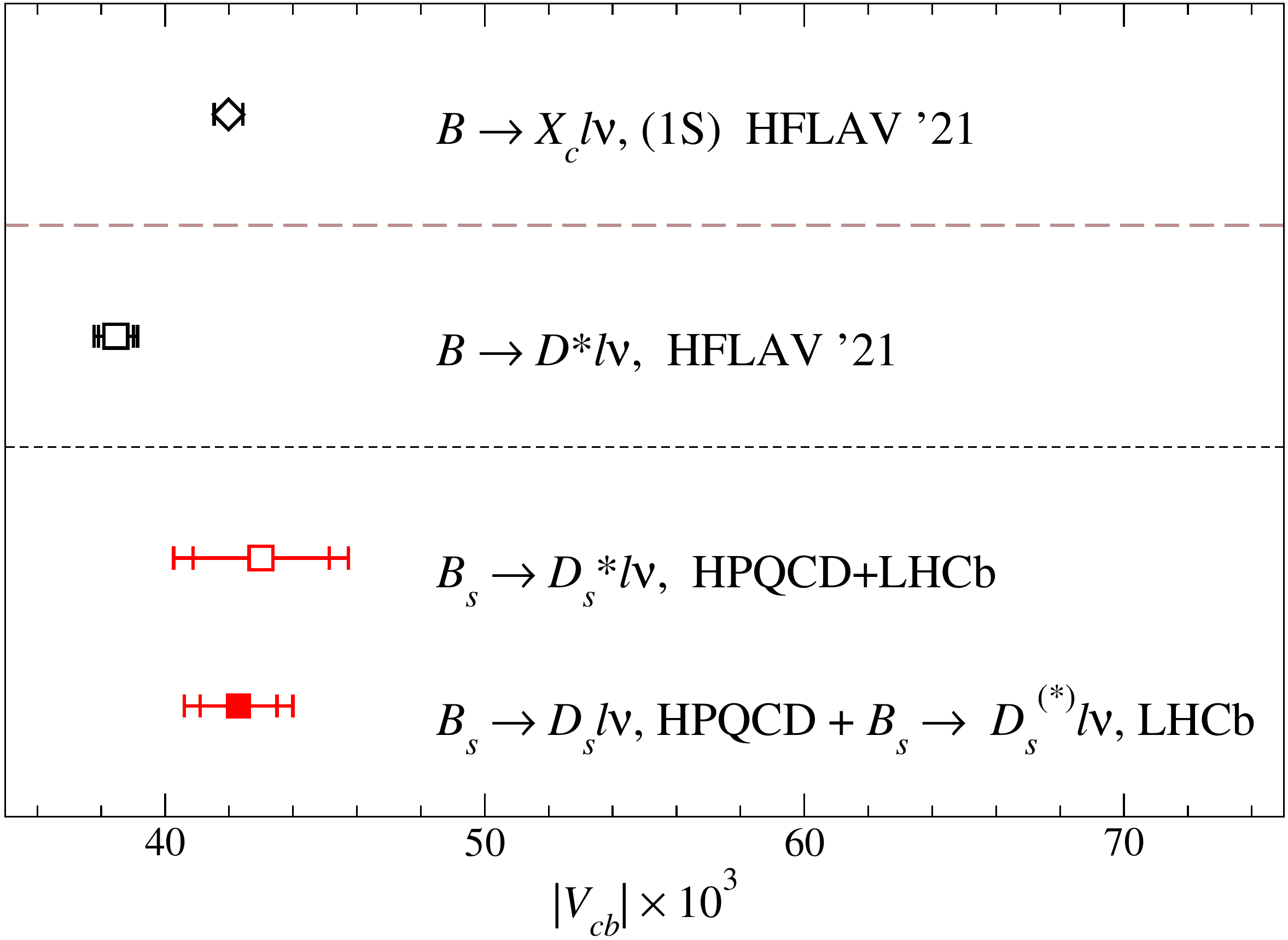}
  \end{minipage}
  \begin{minipage}[h]{0.527\linewidth}                 
    \includegraphics[angle=0,width=1.0\linewidth,clip]%
                    {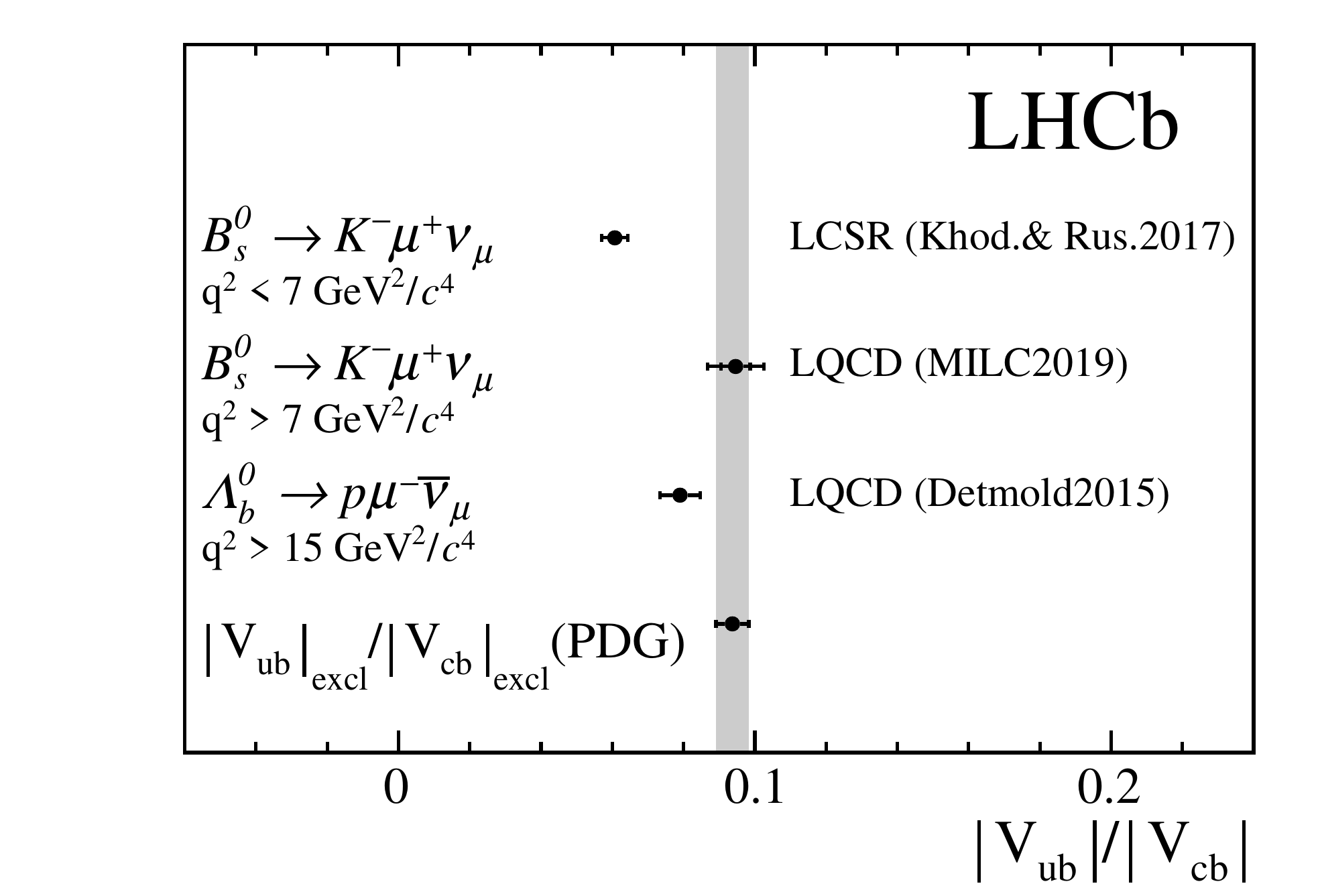}
  \end{minipage}
  \vspace{0mm}
  \caption{
    Left: 
    comparison of $|V_{cb}|$ from semileptonic decays.
    The open red square is obtained by HPQCD's analysis
    of $B_s\!\to\!D_s^*\ell\nu$~\cite{Bs2Dsstar:Nf4:HPQCD:q2full},
    whereas the filled red square is from
    LHCb's analysis~\cite{Bs2Dsstar:LHCb}
    of their data for $B_s\!\to\!D_s^{(*)}\ell\nu$
    with theoretical input of HPQCD's $B_s\!\to\!D_s\ell\nu$
    form factors~\cite{Bs2Ds:Nf4:HPQCD}.
    These are compared with open black symbols 
    obtained by the conventional exclusive ($B\!\to\!D^*\ell\nu$)
    and inclusive analyses\cite{HFLAV22}.
    Right: comparison of recent estimates of $|V_{ub}|/|V_{cb}|$ 
    (figure from Ref.~\cite{Bs2K:LHCb}).
    Two top symbols are from 
    $B_s\!\to\!K\ell\nu$ and $B_s\!\to\!D_s\ell\nu$
    using form factors from QCD light-cone sum rule~\cite{Bs2K:QCDLCSR:KR}
    (top symbol) and from lattice QCD~\cite{Bs2K:Nf3:Fermilab/MILC}
    (second top symbol).
    These are compared with estimates
    from $\Lambda_b\!\to\!p(\Lambda_c)\ell\nu$~\cite{VubVcb:Lambda_b}
    and conventional determination
    from $B\!\to\!\pi\ell\nu$ and $D^{(*)}\ell\nu$~\cite{PDG20}.
  }
  \vspace{-3mm}
  \label{fig:SlD:Vcb}
\end{figure}


The $B_s$ and $B_c$ meson semileptonic decays
including $B_s\!\to\!D_s^{(*)}\ell\nu$, $B_c\!\to\!J/\Psi\ell\nu$
and $B_s\!\to\!K\ell\nu$, provide
independent estimate of $|V_{ub}|$ and $|V_{cb}|$,
as well as the LFUV ratios
$R(X)\!=\!\Gamma(B_{s(c)}\!\to\!X\tau\nu)/\Gamma(B_{s(c)}\!\to\!X\{e,\mu\}\nu)$
as a probe of new physics.
These modes have a couple of advantages on the lattice
over the conventional $B\!\to\!\pi\ell\nu$ and $B\!\to\!D^{(*)}\ell\nu$ decays.
First, they suffer less from statistical fluctuation.
According to Lepage's analysis~\cite{Lepage},
the relative statistical error of the two-point function
exponentially grows as the source-sink separation $\dt$ increases
\bea
   &&
   C_{\bar{Q}q}
   = \langle O_{\bar{Q}q}(\dt) O_{\bar{Q}q}^\dagger(0)\rangle,
   \hspace{5mm}
   \frac{\delta C_{\bar{Q}q}}{C_{\bar{Q}q}}
   \propto
   e^{\alpha \dt},
   \hspace{5mm}
   \alpha = M_{\bar{Q}q} - \frac{M_{\bar{Q}Q}+M_{\bar{q}q}}{2}.
   \label{eqn:SlD:Lepage}
\eea   
The exponent is significantly reduced by changing the spectator quark
from $u,d$ to $s$ or $c$:
for instance,
$\alpha\!\simeq\! 0.31, 0.51, 0.13$ and 0.32 for the $D$, $B$, $D_s$ and $B_s$,
respectively.
In addition,
the chiral extrapolation is expected to be better controlled
thanks to smaller $M_\pi$ dependence without valence pions.
And, in some cases,
the final state hadron with non-zero strangeness or charmness
becomes stable under the strong interaction.


The HPQCD Collaboration has pursued the relativistic calculation
of the $B_s\!\to\!D_s^{(*)}\ell\nu$ form factors with the HISQ heavy quarks~\cite{Bs2Ds:Nf4:HPQCD,Bs2Dsstar:Nf4:HPQCD:q20,Bs2Dsstar:Nf4:HPQCD:q2full}.
As shown in Fig.~\ref{fig:SlD:Vcb},
$B_s\!\to\!D_s^{(*)}\ell\nu$ currently leads to 
4\,--\,6\,\% determination of $|V_{cb}|$,
which is roughly three times worse than $B\!\to\!D^{(*)}\ell\nu$,
and is consistent with both the conventional exclusive and inclusive analyses.
We can, however, expect future improvement on both
theory side (higher statistics on finer lattices)
and experiment side (more data from LHCb Run-II and later).
HPQCD also calculated the LFUV ratio for $B_c\!\to\!J/\Psi\ell\nu$
as $R(J/\Psi)\!=\!0.258(4)$~\cite{Bc2JPsi:Nf4:HPQCD,Bc2JPsi:Nf4:HPQCD:R}.
This is consistent with the LHCb measurement
$R(J/\Psi)\!=\!0.71(25)$~\cite{Bc2JPsi:LHCb}.
The experimental uncertainty is expected to be largely reduced
to $\delta R(J/\Psi) = 0.02$
by future ``Upgrade II'' of LHCb~\cite{LHCb:UpgradeII}.


\begin{figure}[t]
  \centering
  \includegraphics[angle=0,width=0.340\linewidth,clip]%
                  {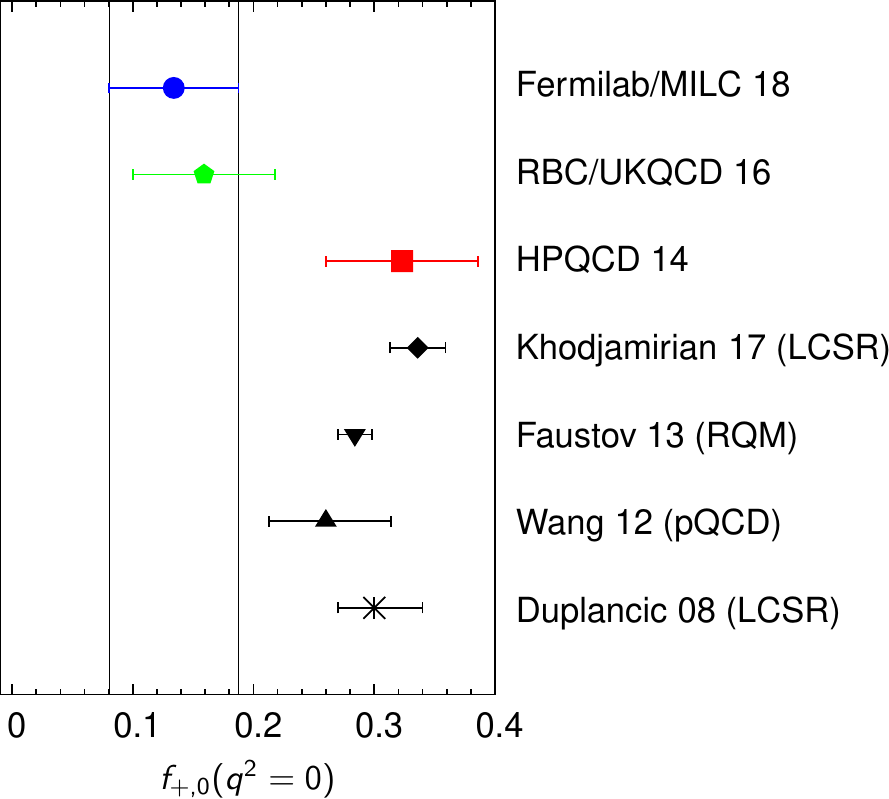}
  \hspace{1mm}
  \includegraphics[angle=0,width=0.640\linewidth,clip]%
                  {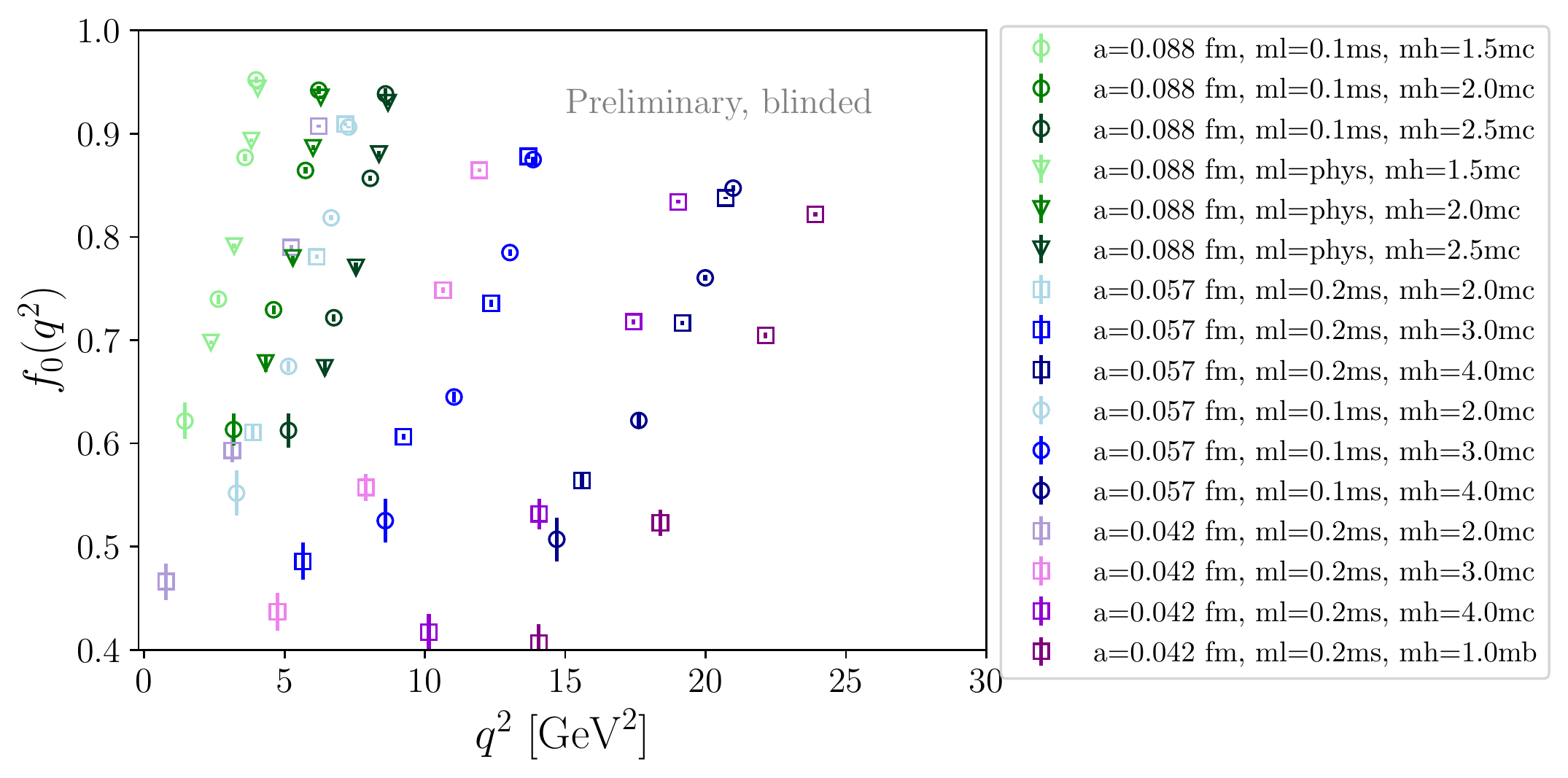}
  \vspace{-2mm}                  
  \caption{
    Left: comparison of $B_s\!\to\!K$ vector and scalar form factor
    at zero momentum transfer $f_+(0)\!=\!f_0(0)$ at the time of 2019
    (figure from Ref.~\cite{Bs2K:Nf3:Fermilab/MILC}).
    The plot includes $f_+(0)$ from
    lattice QCD~\cite{Bs2K:Nf3:HPQCD,B2pi:Nf3:RBC/UKQCD,Bs2K:Nf3:Fermilab/MILC},
    QCD light-cone sum rule (LCSR)~\cite{Bs2K:QCDLCSR:KR,Bs2K:QCDLCSR:D}
    as well as those from relativistic quark model (RQM)~\cite{Bs2K:RQM}
    and NLO perturbative QCD~\cite{Bs2K:pQCD}.
    Right: Fermilab/MILC's preliminary results
    for $B_s\!\to\!K$ scalar form factor $f_0(q^2)$
    as a function of momentum transfer $q^2$
    (figure from Ref.~\cite{B2pi:Nf3:Fermilab/MILC:lat22}).
    Different symbols show data at different lattice spacings,
    light quark masses and bottom quark masses.
    Note that these results are obtained from their blinded analysis.
  }
  \vspace{-3mm}
  \label{fig:SlD:Bs2K}
\end{figure}

The $B_s\!\to\!K\ell\nu$ form factors have been calculated
by HPQCD~\cite{Bs2K:Nf3:HPQCD,Bs2K:Nf3:HPQCD:HISQ},
RBC/UKQCD~\cite{B2pi:Nf3:RBC/UKQCD,BsK:Nf3:RBC/UKQCD}
and Fermilab/MILC~\cite{Bs2K:Nf3:Fermilab/MILC}.
Recently, LHCb observed this CKM suppressed decay~\cite{Bs2K:LHCb},
and estimated $|V_{ub}|/|V_{cb}|$ from the ratio of the branching fractions
$R_{\mathcal B}\!=\!{\mathcal B}(B_s\!\to\!K\ell\nu)/{\mathcal B}(B_s\!\to\!D_s\ell\nu)$.
As shown in the right panel of Fig.~\ref{fig:SlD:Vcb}, however,
there is $\approx\!4\,\sigma$ deviation between two estimates from
different data sets:
$R_{\mathcal B}$ at low $q^2\!<\!7~{\rm GeV}^2$
with form factor input from QCD light-cone sum rule~\cite{Bs2K:QCDLCSR:KR}
and $R_{\mathcal B}$ at high $q^2\!>\!7~{\rm GeV}^2$
with Fermilab/MILC's result for the form factors~\cite{Bs2K:Nf3:Fermilab/MILC}.
This may be attributed to the inconsistency among the form factor results
shown in the left panel of Fig.~\ref{fig:SlD:Bs2K}.
Independent realistic simulations are desired to understand
and resolve the discrepancy.
Fermilab/MILC reported their on-going relativistic simulations
with the HISQ heavy quarks
at the lattice spacing down to $0.04$~fm
and the bottom quark mass $m_b\!\leq\!4 m_c$ close to its physical value~\cite{B2pi:Nf3:Fermilab/MILC:lat22}.
Preliminary result from their blinded analysis is shown in the same figure.


\section{Inclusive semileptonic decays}


Let us consider the $B\!\to\!X_c\ell\nu$ inclusive decay, 
where $X_c$ collectively represents the hadron(s) with single charm quark.
The hadronic tensor $W_{\mu\nu}$ describes
non-perturbative QCD effects to the differential decay rate
\bea
   \frac{d\Gamma}{dq^2 dq^0 dE_\ell}
   & = &
   \frac{G_F^2}{8\pi^3} |V_{cb}|^2 L_{\mu\nu}W^{\mu\nu},
   \label{eqn:incl:ddr}
\eea
where $q^2$ is the momentum transfer to the $\ell\nu$ pair,
$E_\ell$ is the lepton energy in the $B$ rest frame,
and $L_{\mu\nu}$ is the perturbatively calculable leptonic tensor.
Through the optical theorem and operator product expansion (OPE),
$W_{\mu\nu}$ is expressed
as a double expansion in the strong coupling $\alpha_s$
and inverse bottom quark mass $m_b^{-1}$
with $B$ meson matrix elements of local operators
as non-perturbative input
\bea
   W_{\mu\nu}(q)
   & = &
   \sum_{X_c}
   (2\pi)^3 \delta^{(4)}(p-q-r)\frac{1}{2E_B}
   \left\langle B(p) \left| J_\mu^\dagger \right| X_c(r) \right\rangle
   \left\langle X_c(r) \left| J_\nu \right| B(p) \right\rangle
   \nn \\
   & \rightarrow &
   \sum_i
   \frac{C(\alpha_s)}{m_b^{n_i}}
   \left\langle B \left| {\mathcal O}_i \right| B \right\rangle,
   \hspace{10mm}
   \label{eqn:incl:htensor}
\eea
where $J$ is the weak current,
$n_i$ is a positive integer depending on the operator ${\mathcal O}_i$,
and we omit the argument $p$ for $W_{\mu\nu}$ just for simplicity. 
Therefore, the analysis of the inclusive decay has very different systematics
from that of the $B\!\to\!D^{(*)}\ell\nu$ exclusive decay.
Hence comparison of the CKM elements between the exclusive
and inclusive decays provides a rather non-trivial crosscheck,
which has not yet been achieved likely due to
incomplete understanding of systematics.



\begin{figure}[t]
  \centering
  \includegraphics[angle=0,width=0.97\linewidth,clip]%
                  {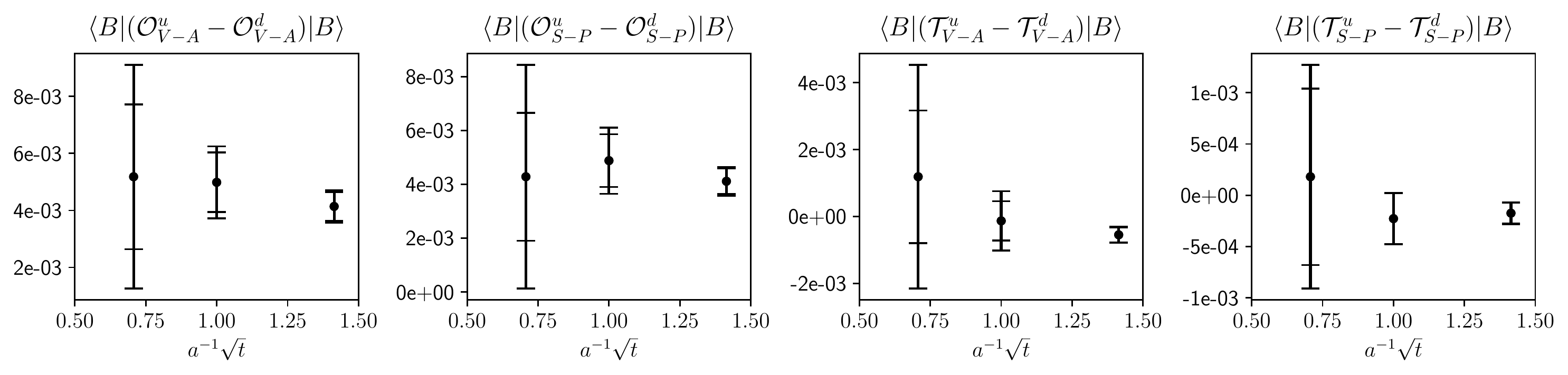}
  \caption{
    Fit result for the bare matrix elements of dimension-6 operators
    as a function of Wilson flow time $\sqrt{t}$
    (figure from Ref.~\cite{HQEdim6:Lin+:Lat22}).
  }
  \vspace{-3mm}
  \label{fig:SlD:SpecEff}
\end{figure}

Lattice QCD can provide first-principles calculation of
the non-perturbative inputs,
such as 
$\mu_\pi^2(\mu)\!=\!(2M_B)^{-1}\langle B|\bar{b}D^2b|B\rangle$ and
$\mu_G^2(\mu)\!=\!(i/4M_B)\langle B|\bar{b}\sigma_{\mu\nu} G_{\mu\nu}b|B\rangle$
for $O(1/m_b^2)$ corrections~\cite{HQEparam:Nf2+:JLQCD}.
Lin reported their calculation of the matrix elements of
the dimension-6 operators 
\bea
   \begin{array}{ll}
     O_{V-A}^q = \left( \bar{b}\gamma_\mu P_L q \right)
                \left( \bar{q}\gamma^\mu P_L b \right),
     &
     O_{S-P}^q = \left( \bar{b} P_L q \right)
                \left( \bar{q} P_R b \right),
     \\
     T_{V-A}^q = \left( \bar{b}\gamma_\mu P_L T^a q \right)
                \left( \bar{q}\gamma^\mu P_L T^a b \right),
     &
     T_{S-P}^q = \left( \bar{b} P_L T^a q \right)
                \left( \bar{q} P_R T^a b \right)
   \end{array}
   \label{eqn:Incl:dim6}
\eea
for $O(1/m_b^3)$ corrections to the heavy quark expansion~\cite{HQEdim6:Lin+:Lat22}.
The matrix elements involving the light valence quark $q$
describe the so-called spectator effects,
which are responsible for the lifetime difference of beauty hadrons,
such as $\tau(B^+)/\tau(B^0)$.
They work in the static limit on RBC/UKQCD gauge ensembles
at two lattice cutoffs $a^{-1}\!\simeq\!1.8$ and 2.4~GeV
and $M_\pi\!\gtrsim\!300$~MeV.
As shown in Fig.~\ref{fig:SlD:SpecEff},
they observe that
both statistical and systematic errors of the fit to extract
the matrix elements from relevant three- and two-point functions
can be largely reduced 
by applying the Wilson flow~\cite{WilsonFlow}
with the flow time $a^{-2}t\!=\!0.5$, 1.0 and 2.0.
Towards the application to the heavy quark expansion,
the renormalization and continuum extrapolation are in progress.


In order to resolve the $|V_{cb}|$ and $|V_{ub}|$ tension, 
a direct lattice calculation of the inclusive decay rate is especially helpful,
since it enables
a detailed comparison between the inclusive and exclusive analyses
in the same simulation.
However, the $B$ meson inclusive decay involves
many non-gold-plated decay channels.
The Lellouch-L\"uscher formalism~\cite{Lellouch-Luescher},
which has been successfully applied to the $K\!\to\!\pi\pi$ decay~\cite{K2pipi:Nf3:RBC/UKQCD:1/2},
becomes increasingly intricate
as the number of the relevant channels increases.
At least currently, it is not straightforward to determine $W_{\mu\nu}(q)$
as a function of $q$.


Recent progress on the inclusive decay is based on an idea that
it would be sufficient to evaluate an integral of $W_{\mu\nu}$
in order to estimate the relevant CKM element.
Let us consider the spectral representation of the lattice four-point function
\bea
  C_{\mu\nu}(t,{\bf q})
  & = &
  \sum_{\bf x}
  \frac{e^{i{\bf q}{\bf x}}}{2M_B} 
  \left\langle B({\bf 0}) \left|
    J_\mu^\dagger({\bf x},t) J_\nu({\bf 0},0)
  \right| B({\bf 0}) \right\rangle
  =
  \int_0^\infty d\omega  W_{\mu\nu}(\omega,{\bf q}) e^{-\omega t},
  \label{eqn:incl:SpecRep}
\eea      
where $\omega\!=\!M_B-q^0$.
It is an ill-posed problem to determine $W_{\mu\nu}(\omega,{\bf q})$
from lattice data of $C_{\mu\nu}(t,{\bf q})$,
because $W_{\mu\nu,L}$ at a finite lattice size $L$
has a largely different functional form
from $W_{\mu\nu}$ on the infinite volume :
the multi-particle continuum part turns into a superposition of
the $\delta$ function like singularities due to the discretized spectrum.
Note also that, practically, only a finite number of discrete data $C_{\mu\nu}$
are available with their statistical error.
The basic idea of Ref.~\cite{SmrSpecFunc:HMR17} is to determine
a smeared spectral function
\bea
  W_{\mu\nu,L,\sigma}(\omega^\prime)
  =
  \int_0^\infty d\omega \Delta_\sigma(\omega^\prime,\omega) W_{\mu\nu,L}(\omega),
  \label{eqn:incl:SmrSpecFunc}
\eea
where the smearing function $\Delta_\sigma(\omega^\prime,\omega)$ is
a smooth approximation of the $\delta$ function:
namely, ${\rm lim}_{\sigma \to 0} \Delta_\sigma(\omega^\prime,\omega)
\!=\! \delta(\omega^\prime-\omega)$.
Estimating the smooth function $W_{\mu\nu,L,\sigma}(\omega^\prime)$
from $C_{\mu\nu}$ is a well-posed problem.
The infinite volume spectral function is given by the double limit
$W_{\mu\nu}(\omega)\!=\!{\rm lim}_{\sigma\to 0}{\rm lim}_{L\to\infty} 
W_{\mu\nu,L,\sigma}(\omega)$,
where the order of the limits is not commutable.
We refer the reader to the review talk by Bulava
for more details and interesting applications of this approach~\cite{SpecFunc:Bulava:Lat22}.


In Ref.~\cite{EnergyInt:GH20},
Gambino and Hashimoto proposed a method,
which is more directly applicable to the inclusive processes.
Instead of the smearing function $\Delta_\sigma(\omega^\prime,\omega)$,
the $\omega$-integral kernel dictated by the kinematical factor and
leptonic tensor in Eq.~(\ref{eqn:incl:ddr})
is used to directly evaluate the inclusive rate
\bea
  \Gamma
  & = &
  \frac{G_F^2}{24\pi^3} |V_{cb}|^2
  \int_0^{{\bf q}^2_{\rm max}} d{\bf q}^2 \sqrt{{\bf q}^2} \bar{X}({\bf q}^2),
  \hspace{3mm}
  \bar{X}_{L,\sigma}({\bf q}^2)
  = \int_{\omega_{\rm min}}^{\omega_{\rm max}} d\omega
    K_{\mu\nu,\sigma}(\omega, {\bf q}) W_{\mu\nu,L}(\omega,{\bf q}),
  \hspace{5mm}
  \label{eqn:incl:DR:GH}
\eea
where $w_{\rm min}\!=\!\sqrt{M_D^2+{\bf q}^2}$
and $w_{\rm max}\!=\!M_B - \sqrt{{\bf q}^2}$.
Note that $K_{\mu\nu}$ involves the Heaviside step function 
$\theta(\omega_{\rm max}-\omega)$
to realize the upper limit of the $\omega$-integral.
This is approximated by a smooth function,
for instance a sigmoid function $\theta_\sigma(x)\!=\!1/(1+e^{-x/\sigma})$,
in $K_{\mu\nu,\sigma}$.
Therefore, the double limit $\bar{X}({\bf q}^2)
\!=\!{\rm lim}_{\sigma\to 0}{\rm lim}_{L\to\infty} \bar{X}_{L,\sigma}({\bf q}^2)$
must be taken also in this method.
We note that, as discussed in Ref.\cite{EnergyInt:FHKO},
this method can be applied to other inclusive processes,
such as the charged current neutrino-nucleon scattering
important for the neutrino experiments,
by appropriately choosing the integral kernel.


\begin{figure}[t]
  \centering
  \begin{minipage}[h]{0.507\linewidth}                 
    \includegraphics[angle=0,width=1.0\linewidth,clip]%
                    {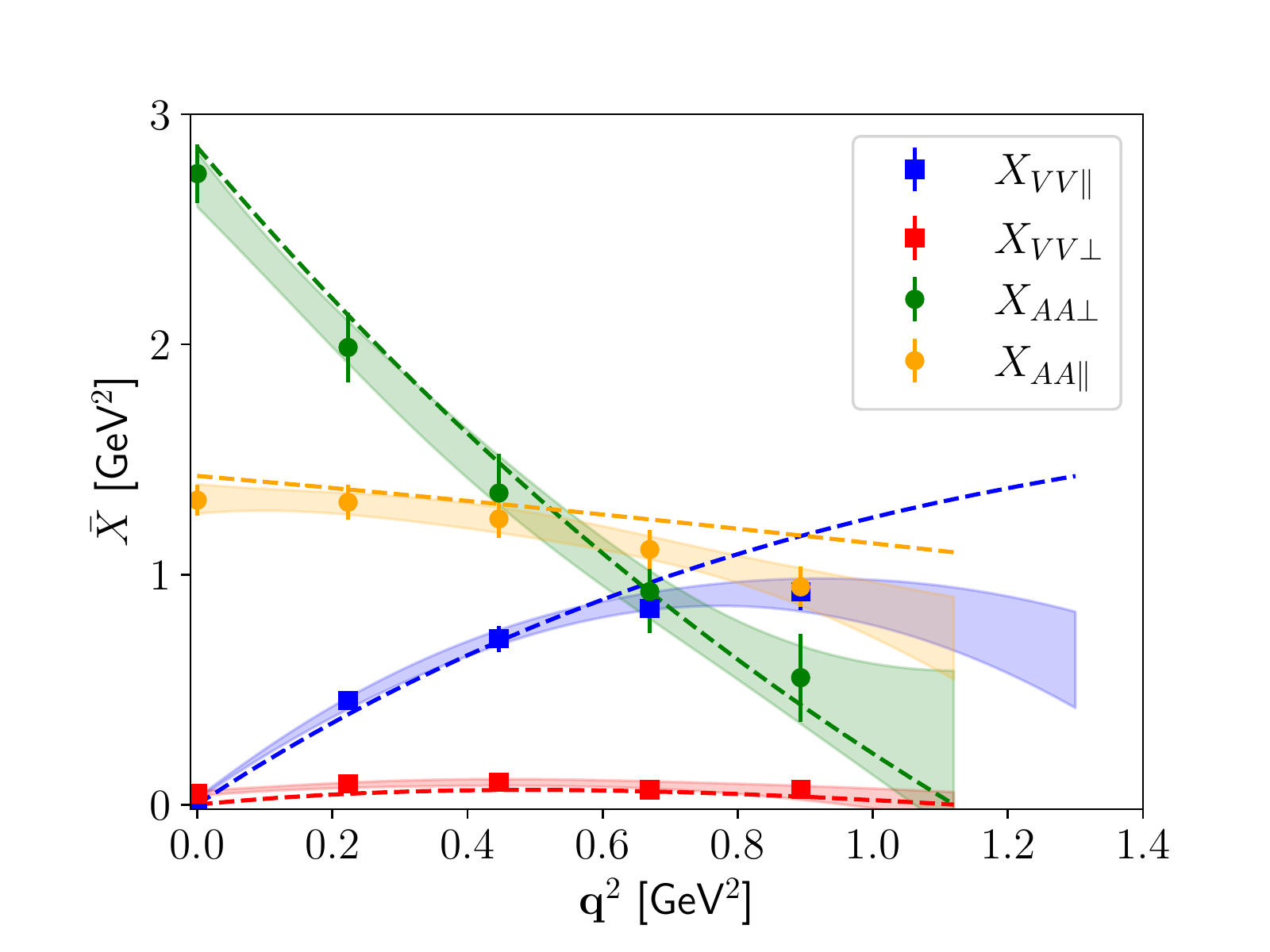}
  \end{minipage}
  \begin{minipage}[h]{0.487\linewidth}                 
    \includegraphics[angle=0,width=1.0\linewidth,clip]%
                    {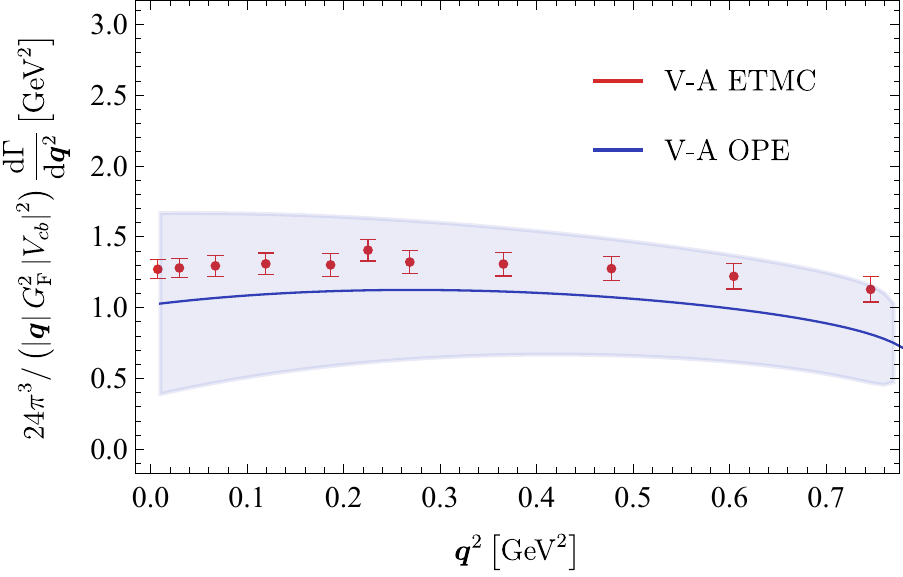}
  \end{minipage}
  \vspace{-2mm}
  \caption{
    Left:
    comparison of $\bar{X}_{L,\sigma\to 0}({\bf q}^2)$ 
    with that from exclusive decay.
    Symbols and bands show $\bar{X}_{L,\sigma}({\bf q}^2)$
    in Eq.~(\ref{eqn:incl:DR:GH})
    for the $B_s\!\to\!X_{sc}\ell\nu$ inclusive decay
    in the $\sigma\!\to\!0$ limit.
    Different symbols show data with different choices
    of the weak currents (vector or axial vector)
    and their polarization (perpendicular or not-perpendicular to ${\bf q}$)
    for $C_{\mu\nu}$.
    These are compared with the ground state contribution
    from the $B_s\!\to\!D_s^{(*)}\ell\nu$ exclusive decay (dashed lines).
    All data are calculated on a JLQCD ensemble at $a^{-1}\!\approx\!3.6$~GeV
    and $M_\pi\!\approx\!300$~MeV~\cite{B2pi:Nf3:JLQCD}.
    Right: 
    comparison of $\bar{X}_{L,\sigma\to 0}({\bf q}^2)$ 
    with conventional OPE calculation.
    The red circles are calculated
    on an ETM ensemble at $a^{-1}\!\approx\!2.5$~GeV
    and $M_\pi\!\approx\!400$~MeV ~\cite{Conf:Nf4:ETM}.
    The OPE prediction includes power corrections up to and including
    $O(1/m_b^{2,3})$ and $O(\alpha_s)$.
    (The left and right panels are taken from Ref.~\cite{Incl:Nf3:JLQCD+ETM+}.)
  }
  \vspace{-3mm}
  \label{fig:incl:gambino+}
\end{figure}

At this conference,
Smecca~\cite{Incl:Nf3:JLQCD+ETM+:Lat22} reported on
their feasibility study of this method~\cite{Incl:Nf3:JLQCD+ETM+}.
They studied
the computationally inexpensive $B_s\!\to\!X_{cs}\ell\nu$ inclusive decay,
and made a detailed comparison among the conventional OPE calculation
and lattice results 
obtained on ETM and JLQCD ensembles.
The left panel of Fig.~\ref{fig:incl:gambino+} compares the JLQCD data of 
$\bar{X}_{L,\sigma}({\bf q}^2)\!=\!(24\pi^3/G_F^2|V_{cb}|^2|{\bf q}|)(d\Gamma/d{\bf q}^2)$
appearing in Eq.~(\ref{eqn:incl:DR:GH}) in the $\sigma\!\to\!0$ limit.
They observe good consistency between the full inclusive contribution
of $B_s\!\to\!X_{cs}\ell\nu$ and its ground state contribution
from the $B_s\!\to\!D_s^{(*)}\ell\nu$ exclusive decay.
Note that these JLQCD data are obtained by using
relativistic domain-wall bottom quarks
with an unphysically small $m_b\!\simeq\!2.44 m_c$
corresponding to $M_{B_s}\!\sim\!3.5$~GeV.
The authors discussed that
the ground state saturation can be expected
from the limited phase space as well as heavy quark symmetry
due to the small $m_b$ close to the charm mass $m_c$.
The good consistency, therefore, demonstrates
the validity of the inclusive analysis
with very different systematics
(for instance, the use of the $B_s\!\to\!B_s$ four-point function $C_{\mu\nu}$,
and the approximated kernel $K_{\mu\nu,L,\sigma}$)
from the exclusive analysis.


\begin{table}[t]
  \centering
  \small
  \caption{
    Total width divided by $|V_{cb}|^2$
    obtained from OPE and lattice calculations
    (results from Ref.~\cite{Incl:Nf3:JLQCD+ETM+}).
  }
  \vspace{-2mm}
  \label{tbl:sim:param}
  \begin{tabular}{l|ll|ll}
    \hline
    $m_b/m_c$ & \multicolumn{2}{c|}{$\sim$~2.4} & \multicolumn{2}{c}{$\sim$~2.0}
    \\ \hline
              & JLQCD    & OPE      & ETM       & OPE
    \\ \hline
    $\Gamma/|V_{cb}|^2 \times 10^{13}$ [GeV] 
              & 4.46(21) & 5.7(9)   & 0.987(60) & 1.20(46)
    \\ \hline
  \end{tabular}
  \label{tbl:incl:rate}
\end{table}

In the right panel of Fig.~\ref{fig:incl:gambino+},
on the other hand,
the ETM data of $\bar{X}_{L,\sigma\!\to 0}({\bf q}^2)$
are compared with the conventional OPE calculation
including $O(1/m_b^3)$ and $O(\alpha_s)$ corrections.
The ETM data are also obtained at unphysically small $m_b\!\approx\!2 m_c$,
but note that the comparison with the OPE can be made
directly at the simulated $m_b$.
They observe a reasonable consistency in $\bar{X}_{L,\sigma\to 0}({\bf q}^2)$,
which persists to the total rate $\Gamma/|V_{cb}|^2$
after integrating over ${\bf q}^2$ 
as shown in Table~\ref{tbl:incl:rate}.
%
%
However, they also observe tensions
between lattice and OPE calculations 
through the decomposition into the vector and axial vector components
as well as the perpendicular and longitudinal polarization components
of $\bar{X}_{L,\sigma\to 0}({\bf q}^2)$.
This could be a manifestation of the violation
of the quark-hadron duality,
and hence deserves more detailed investigation in the future.


To that end, we need to more carefully study the systematics of this approach.
As mentioned,
the step function $\theta(x)$
in the $\omega$-integration kernel $K_{\mu\nu}$ in Eq.~(\ref{eqn:incl:DR:GH})
is approximated by a smooth function
$\theta_\sigma(x)$, which goes to $\theta(x)$ in the $\sigma\!\to\!0$ limit.
In order to evaluate the $\omega$-integral using lattice data $C_{\mu\nu}(t)$,
we further approximate the smooth kernel $K_{\mu\nu,\sigma}$
with polynomials in $e^{-\omega}$~\cite{EnergyInt:GH20}.
To this end, there have been two proposals.
References~\cite{SmrSpecFunc:HMR17,SmrSpecFunc:HLT19}
proposed to use the so-called Backus-Gilbert method~\cite{BGapprox:'68,BGapprox:'70}.
It approximates $K_{\mu\nu,\sigma}$ with the basis functions $b_t(\omega)\!=\!e^{-\omega t}$ as 
\bea
  K_{\mu\nu,\sigma}
  & \simeq &
  \sum_{t}^{N} g_t b_t(\omega),
  \label{eqn:incl:bg:approx}
\eea  
where the coefficients $\{g_t\}$ are determined by minimizing the norm
\bea
  \int_{\omega_0}^{\infty} d\omega \left|K_{\mu\nu,\sigma}-\sum_t^N g_t b_t(\omega)\right|^2,
  \label{eqn:incl:bg:meas}
\eea
as long as the polynomial approximation is concerned.
Another proposal~\cite{EnergyInt:GH20} employs
the (shifted) Chebyshev polynomial $T_k^*(e^{-\omega t})$ as
\bea
  K_{\mu\nu,\sigma}
  & \simeq &
  \frac{c_0^*}{2} + \sum_k^{N} c_k^* T_k^*(e^{-\omega t}),
  \label{eqn:incl:cheby}
\eea  
where the maximum deviation is minimized (the min-max approximation).
We note that $t$ in these approximations is identified with
the argument of $C_{\mu\nu}(t)$ to evaluate the $\omega$-integral.

\begin{figure}[t]
  \centering
  \begin{minipage}[h]{0.507\linewidth}                 
    \includegraphics[angle=0,width=1.0\linewidth,clip]%
                    {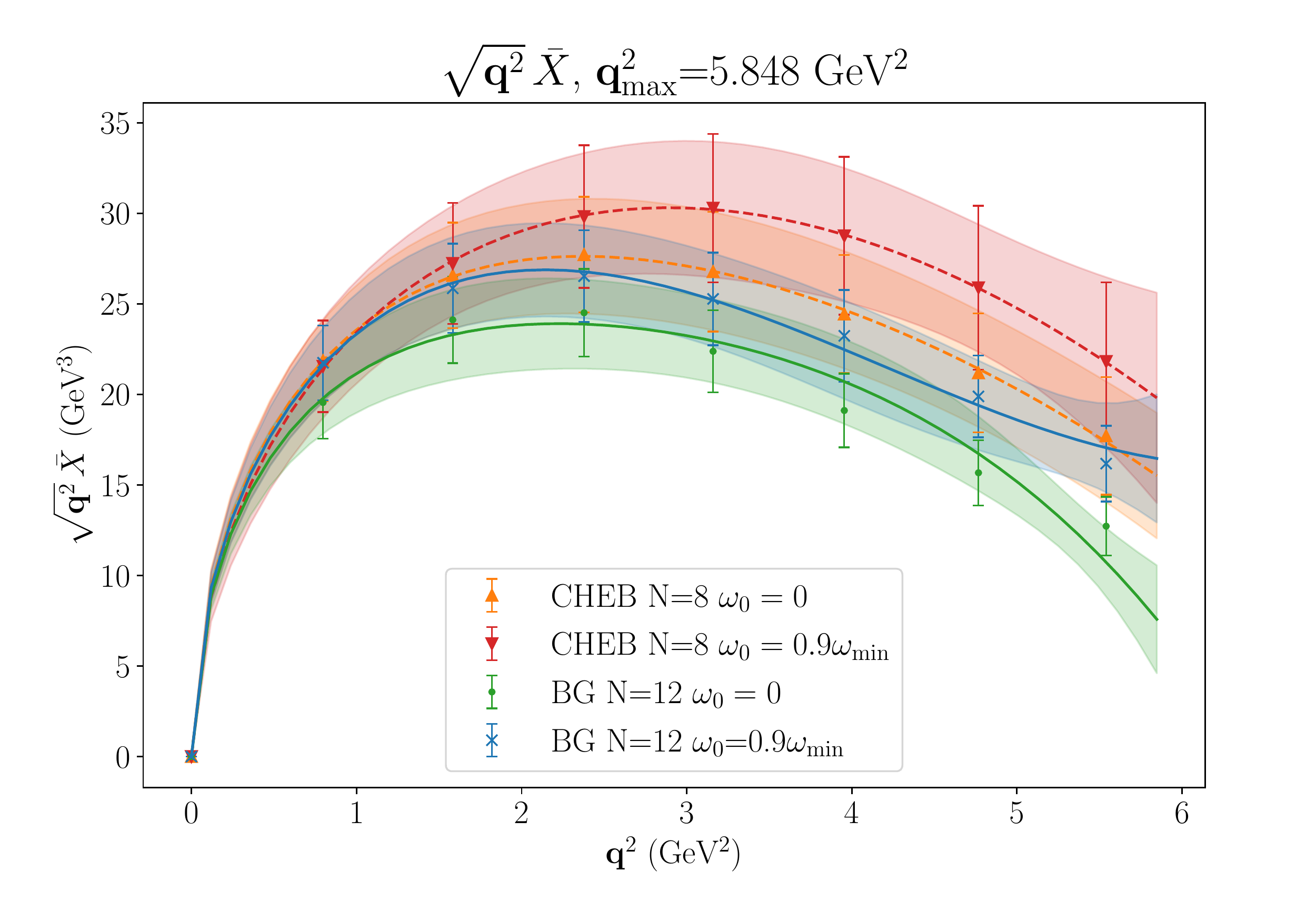}
  \end{minipage}
  \begin{minipage}[h]{0.487\linewidth}                 
    \includegraphics[angle=0,width=1.0\linewidth,clip]%
                    {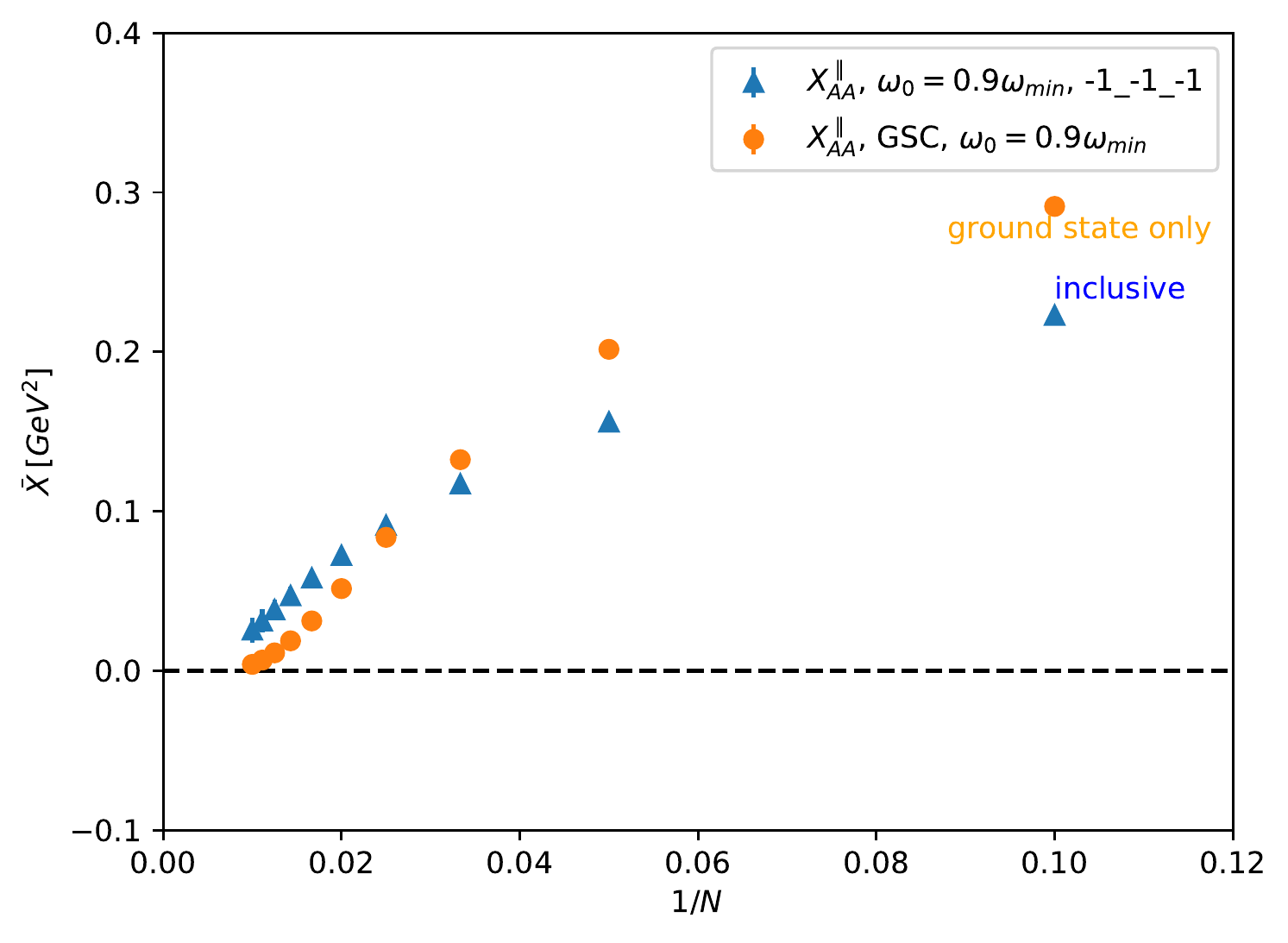}
  \end{minipage}
  \vspace{-2mm}
  \caption{
    Left:
    comparison of $\sqrt{{\bf q}^2} \bar{X}_{L,\sigma}({\bf q}^2)$
    between Backus-Gilbert method and Chebyshev approximation
    as well as different values of $\omega_0$
    (figure from Ref.~\cite{Incl:Nf3:RBC/UKQCD}).
    All data are calculated on a RBC/UKQCD ensemble
    at $a^{-1}\!\approx\!1.8$~GeV and $M_\pi\!\approx\!330$~MeV.
    Right:
    axial vector component $X_{AA,L,\sigma}^{\parallel}({\bf q}^2)$ of
    $\bar{X}_{L,\sigma}({\bf q}^2)$
    as a function of $1/N = \sigma$
    (figure from Ref.~\cite{Incl:Nf3:JLQCD}).
    The blue triangles show the full inclusive result,
    whereas the orange circles are its ground state contribution.
  }
  \vspace{-3mm}
  \label{fig:incl:rbc/ukqcd+jlqcd}
\end{figure}


Barone {\it et al.} carried out a systematic study
on this approximation step~\cite{Incl:Nf3:RBC/UKQCD}.
The $\omega$ integral $\sqrt{{\bf q}^2}\bar{X}_{L,\sigma}({\bf q}^2)$
is evaluated by using the Backus-Gilbert and Chebyshev approaches
as well as by varying $\omega_0$,
which is the lower cut of the $\omega$-region to approximate $K_{\mu\nu,\sigma}$.
They employ the relativistic heavy quark (RHQ) action~\cite{RHQ}
based on HQET to simulate the physical mass $\mbphys$
on a RBC/UKQCD ensemble at $a^{-1}\!\simeq\!1.8$~GeV.
The reasonable consistency shown
in the left panel of Fig.~\ref{fig:incl:rbc/ukqcd+jlqcd}
demonstrates
the stability of the inclusive analysis
against the choice of the approximation method and $\omega$-region thereof.
We also note that their simulation at $\mbphys$ may observe
significant contribution of the excited states to the inclusive rate.


The same authors also studied another important systematics:
namely, the extrapolation to the $\sigma\!\to\!0$ limit
and the associated uncertainty.
As reported by Kellermann~\cite{Incl:Nf3:JLQCD},
on the JLQCD ensemble mentioned above,
they focus on the $D_s\!\to\!X_{ss}\ell\nu$ inclusive decay,
which can be precisely studied even through the relativistic approach,
because all relevant valence quark masses can be set to their physical value
with discretization effects under control.
They observe that the $\sigma$ extrapolation is essential to quantitatively
estimate the differential decay rate.
The right panel of Fig.~\ref{fig:incl:rbc/ukqcd+jlqcd}
shows the axial current component $X_{AA,L,\sigma}^{\parallel}$ 
with their polarization non-perpendicular to ${\bf q}$.
This receives
the leading contribution from the $s\bar{s}$-vector state $\varphi$,
since $\langle \eta_s | A_\mu | B \rangle$ vanishes
due to parity symmetry.
However, this also vanishes 
with the chosen momentum ${\bf q}\!=\!(1,1,1)$ (in units of $2\pi/L$),
because $E_{\varphi} \!>\! M_{D_s}$.
The figure shows that $X_{AA}^\parallel$ is not zero at large values of
$\sigma\!=\!1/N$ due to the error of the kernel approximation,
and it approaches zero at $\sigma \!\lesssim\! 0.01$
suggesting the importance of the $\sigma$ extrapolation.

They also provided an error estimate
based on a mathematical property of the Chebyshev polynomial
$|T_k^*(e^{-\omega t})|\!\leq\!1$,
namely by enforcing the mathematical upper and lower limits of $T_k^*$.
This covers the ground state contribution though being too conservative,
and further study for more realistic error estimate is needed.
Another important issue is to approach the infinite volume limit
$L\!\to\!\infty$,
which has not been studied for the application to heavy meson inclusive decays.


\section{Leptonic decays}


\begin{figure}[t]
  \centering
  \begin{minipage}[h]{0.532\linewidth}                 
    \includegraphics[angle=0,width=1.0\linewidth,clip]%
                    {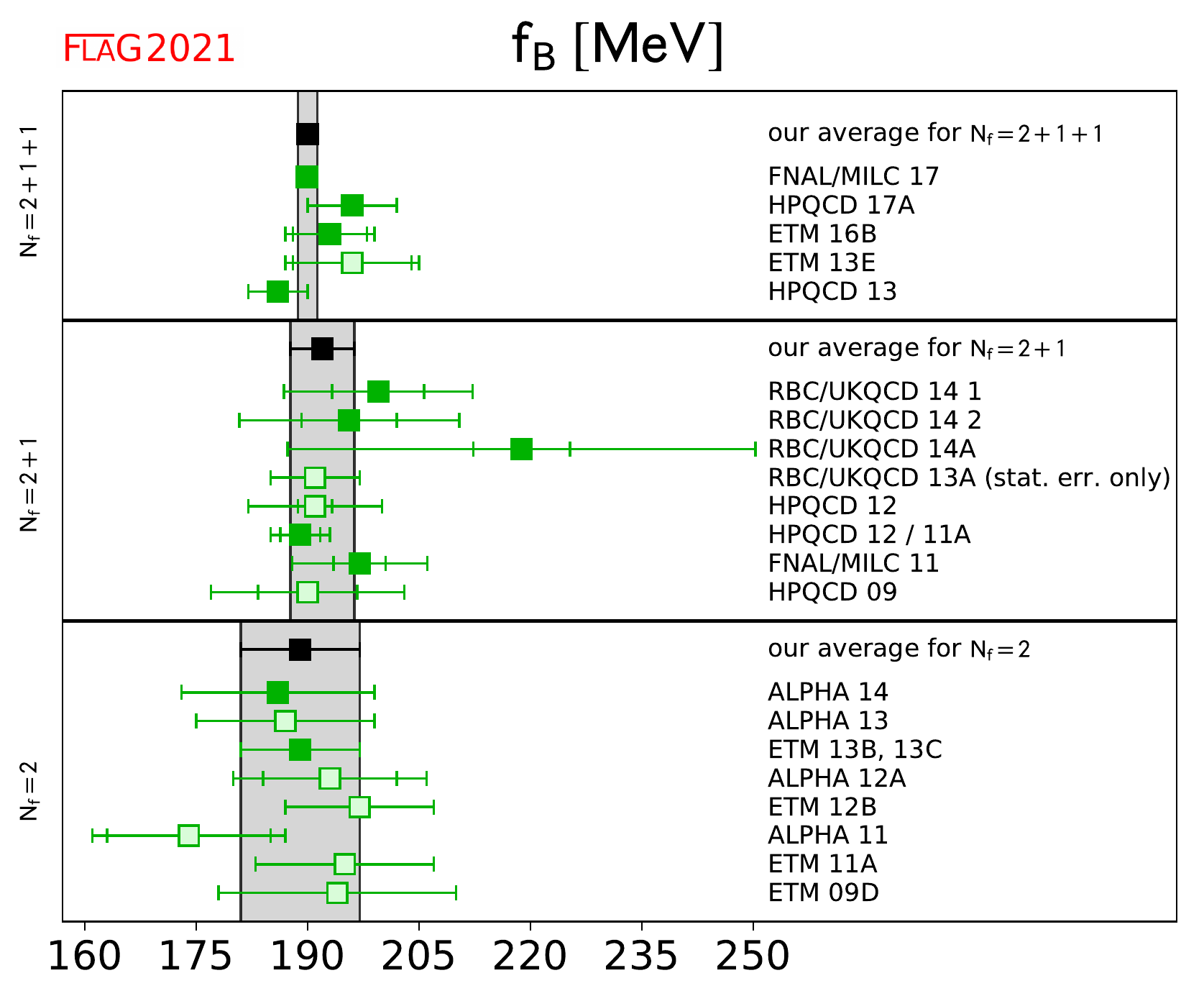}
  \end{minipage}
  \hspace{5mm}
  \begin{minipage}[h]{0.422\linewidth}                 
    \includegraphics[angle=0,width=1.0\linewidth,clip]%
                    {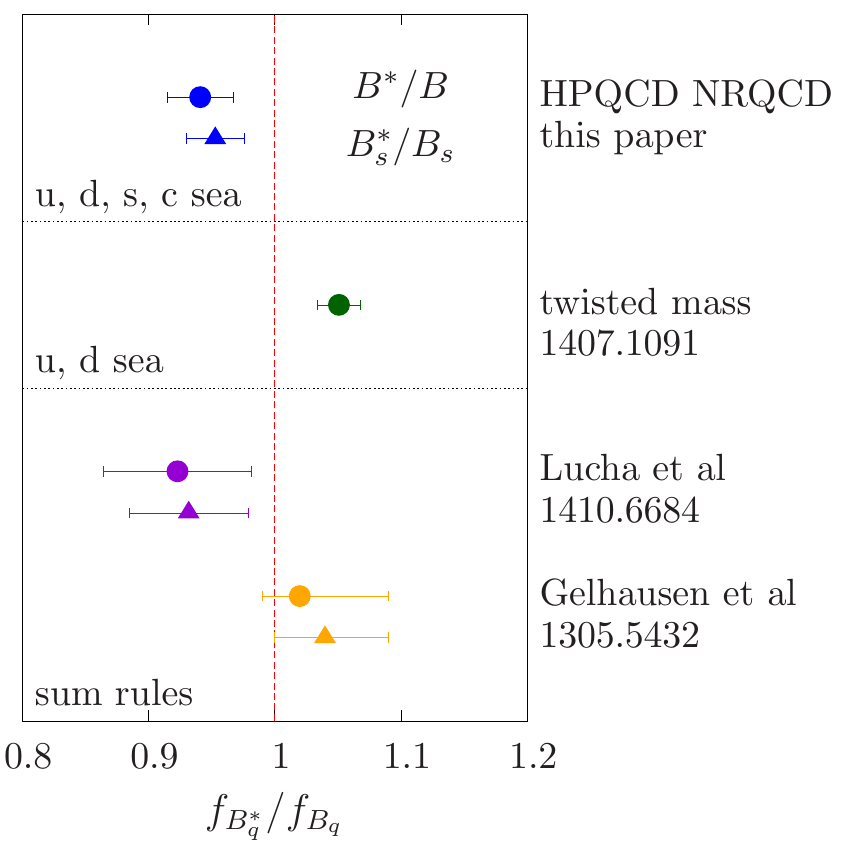}
  \end{minipage}
  \vspace{-2mm}
  \caption{
    Left:
    comparison of realistic calculations of $f_B$
    (figure from the latest FLAG review~\cite{FLAG5}).
    The green squares are individual studies,
    whereas the black squares and bands represent their average
    for $N_f\!=\!2+1+1$ 2+1 and 2.
    Right:
    previous estimates of $f_{B^*}/f_B$
    (figure from Ref.~\cite{fB*:Nf3:HPQCD}).
    The blue~\cite{fB*:Nf3:HPQCD} and black~\cite{fB*:Nf3:BYORS} symbols
    show results from lattice QCD,
    whereas the purple~\cite{fB*:QCDSR:LMS} and orange~\cite{fB*:QCDSR:GKPR}
    symbols are from QCD sum rules.
  }
  \vspace{-3mm}
  \label{fig:LD:fB*}
\end{figure}

The $B\!\to\ell\nu$ leptonic decay provides
an alternative determination of $|V_{ub}|$
as well as interesting probe of new physics
sensitive only to odd parity interactions.
Through many independent calculations
shown in the left panel of Fig.~\ref{fig:LD:fB*},
the relevant hadronic input, namely the decay constant $f_B$,
is determined with the $\approx\!0.7$\,\% accuracy,
at which the electromagnetic (EM) corrections would be no longer negligible.
The accuracy of $|V_{ub}|\!=\!4.05(3)_{\rm th}(64)_{\rm exp}$~\cite{FLAG5} is
largely limited by the statistics of experimental data
due to the helicity suppression.
B2TiP expects that
the experimental uncertainty will be largely reduced to a few \%
at the target luminosity of Belle II,
and $B\!\to\!\ell\nu$ will determine $|V_{ub}|$
with an accuracy competitive to $B\!\to\!\pi\ell\nu$.


The decay constant $f_{B_s}$ of the neutral $B_s$ meson
describes the $B_s\!\to\!\mu\mu$ decay,
which is mediated by the flavor changing neutral current (FCNC),
and hence is expected to be sensitive to new physics.
Previously, a combined analysis of 
the branching fraction ${\mathcal B}(B_s\!\to\!\mu\mu)$
by the ATLAS, CMS and LHCb experiments
reported $\sim\!2.5\,\sigma$ tension
with the SM prediction~\cite{Bs2mumu:ATLAS+CMS+LHCb:2020},
which, however, became insignificant in the latest CMS result~\cite{Bs2mumu:CMS:ICHEP2022}.
Similar to $f_B$,
many independent lattice calculations
leads to the FLAG average of 0.6\,\% accuracy
leading to the theoretical precision of
$\Delta{\mathcal B}(B_s\!\to\!\mu\mu)\!\sim\!4$\,\%.
The current experimental accuracy of $\approx\!10$\,\% is lower than the SM prediction,
but is expected to become competitive by HL-LHC.


\begin{figure}[b]
  \centering
  \begin{minipage}[h]{0.452\linewidth}                 
    \includegraphics[angle=0,width=1.0\linewidth,clip]%
                    {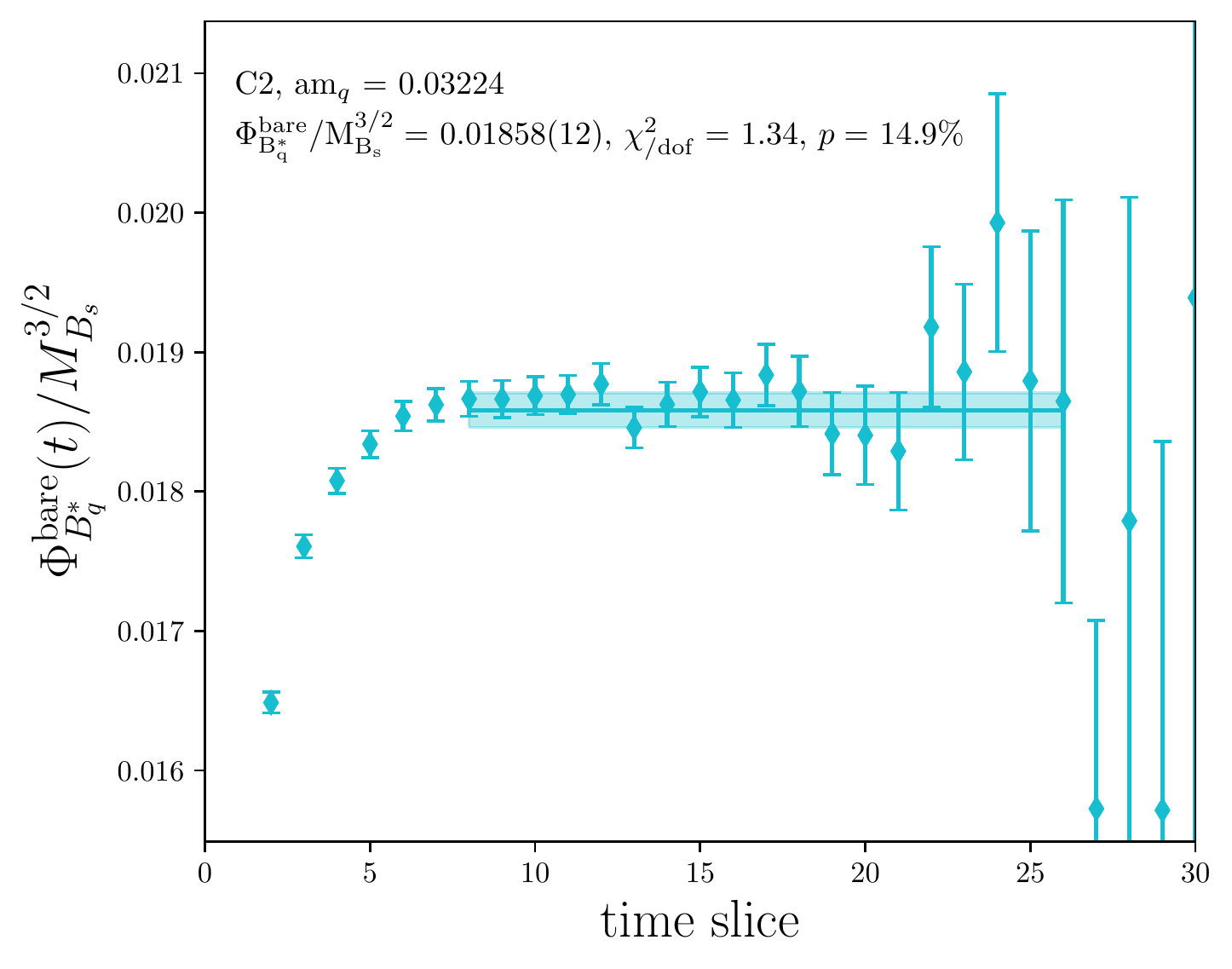}
  \end{minipage}
  \begin{minipage}[h]{0.542\linewidth}                 
    \includegraphics[angle=0,width=1.0\linewidth,clip]%
                    {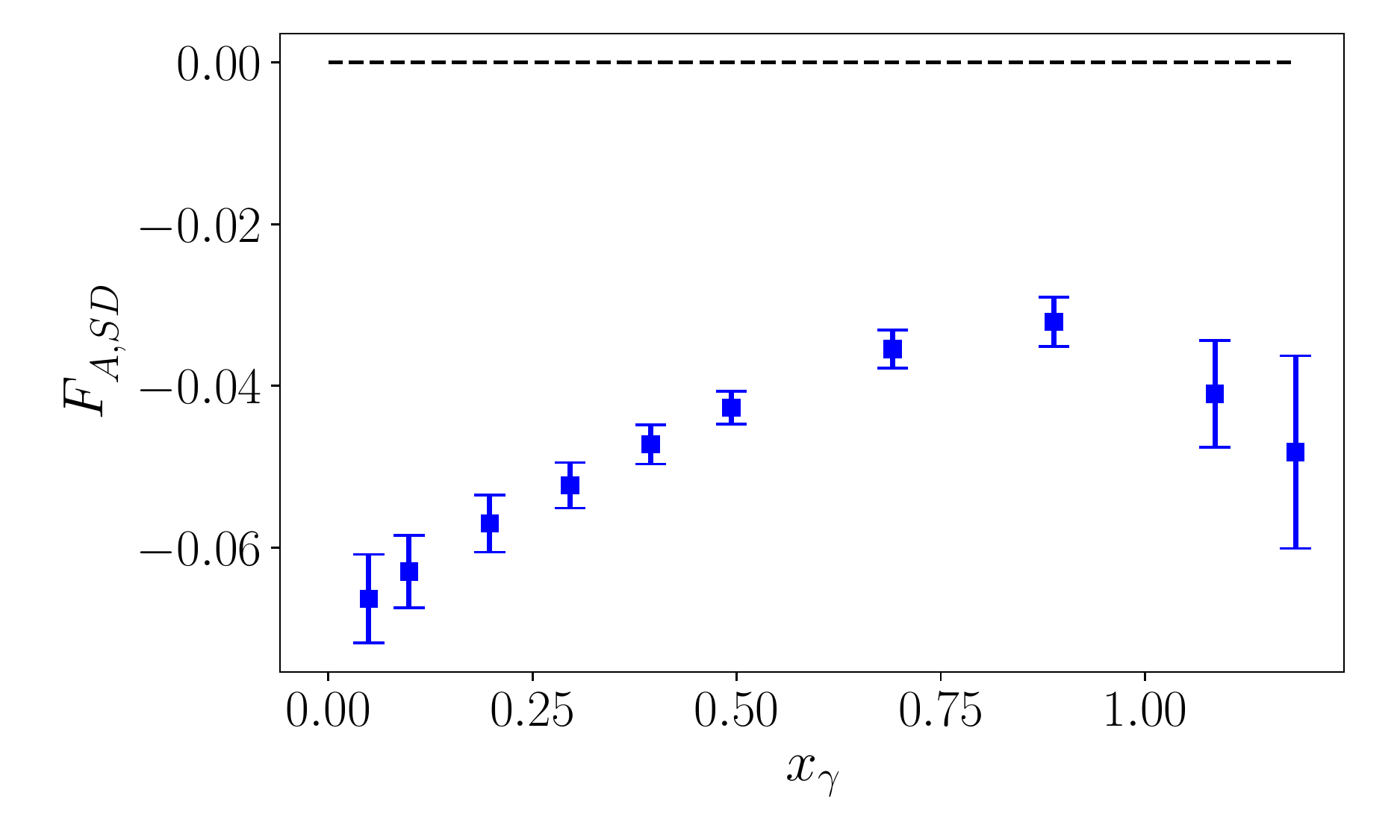}
  \end{minipage}
  \vspace{-2mm}
  \caption{
    Left:
    effective plot of dimensionless decay amplitude 
    $\Phi_{B^*}/M_{B_s}^{3/2}\!=\!f_{B^*}\sqrt{M_{B^*}}/M_{B_s}^{3/2}$
    (figure from Ref.~\cite{fB*:RBC/UKQCD:Lat22}.)
    Right:
    structure dependent part of axial vector form factor $F_{A,\rm SD}$
    as a function of $x_\gamma\!=\!2 v p_\gamma/M_P$
    (figure from authors of Ref.~\cite{P2ellnugamma:Nf3:Giusti+,P2ellnugamma:Nf3:Giusti+:paper}).
  }
  \vspace{-3mm}
  \label{fig:Ld:fB*+rad}
\end{figure}

Having achieved the good precision for $f_{B_{(s)}}$,
it is interesting to explore different observables.
One direction is to extend to other mesons, such as $B^*$ and $B_c$.
In contrast to $f_{B_{(s)}}$, however,
there have been much less studies of the vector meson decay constant $f_{B^*}$
as shown in the right panel of Fig.~\ref{fig:LD:fB*}.
Due to the tension between the lattice studies,
more independent studies are highly welcome.
At this conference, RBC/UKQCD reported on their on-going study
of $f_{B^*}$ as well as $f_{B_c}$~\cite{fB*:RBC/UKQCD:Lat22}.
They simulate four lattice cutoffs up to $a^{-1}\!\sim\!3.2$~GeV
and pion masses down to $M_\pi\!\sim\!270$~MeV
with the RHQ action for bottom quarks.
While the continuum extrapolation has to be done,
they observe reasonable signal and ground-state saturation
for both the $B^*$ mass and dimensionless decay amplitude
$\Phi_{B^*}/M_{B_s}^{3/2}\!=\!f_{B^*}\sqrt{M_{B^*}}/M_{B_s}^{3/2}$
as shown in the left panel of Fig.~\ref{fig:Ld:fB*+rad}.


Another interesting direction is to study the radiative leptonic decay
$B\!\to\!\ell\nu\gamma$.
This mode potentially provides an alternative determination of $|V_{ub}|$,
since the helicity suppression is lifted with a photon in the final state,
and Belle II is expected to measure the branching fraction
${\mathcal B}(B\!\to\!\ell\nu\gamma)$ with 4\,\% accuracy~\cite{B2TiP}.
For large photon energy $E_\gamma$,
this would be a clean probe of the internal structure of the $B$ meson,
for example the first inverse moment $1/\lambda_B$ of the light-cone
distribution amplitude.
The RM123+SOTON collaboration has studied
the radiative decays $P\!\to\!\ell\nu\gamma$
of light mesons ($P\!=\!\pi, K$) in the full region of the photon energy,
and $D_{(s)}\!\to\!\ell\nu\gamma$ for soft photons ($E_\gamma\!\leq\!400$~MeV).
They employ the twisted mass discretization and simulate
four lattice cutoffs up to $a^{-1}\!\leq\!3.2$~GeV
and pion masses down to $M_\pi\!=\!220$~MeV.


At this conference, Giusti reported their study of
the $K$ and $D_{(s)}$ radiative decays in full photon energy region
aiming at an extension to
$B\!\to\!\ell\nu\gamma$~\cite{P2ellnugamma:Nf3:Giusti+}.
The relevant correlators are calculated on a RBC/UKQCD ensemble
at $a^{-1}\!\sim\!1,8$~GeV and $M_\pi\!\simeq\!340$~MeV.
The matrix element is decomposed as 
\bea
   &&
   -i \int d^4x e^{ip_\gamma x}
   \left\langle 0 \left|
      T\left[ J_\mu^{({\rm EM})}(x) J_\nu^{({\rm weak})}(0) \right]
   \right| P(p_P) \right\rangle
   \nn \\
   & = &
   \varepsilon_{\mu\nu\rho\sigma} p_\gamma^{\rho} v^\sigma F_V
 + i\left[ -g_{\mu\nu} (v p_\gamma) + v_\mu (p_\gamma)_\nu \right] F_A
 - i\frac{v_\mu v_\nu}{v p_\gamma}M_P f_P
 + (p_\gamma)_\mu{\rm -terms},
   \label{eqn:RLD:FFs}
\eea
where
$F_A$ ($F_V$) represents the axial (vector) form factor,
and $f_P$ is the decay constant.
The last term represents the contributions proportional to $(p_\gamma)_\mu$,
which vanishes when contracted with $p_\gamma$.
For large photon energy,
the decay amplitude is described by $F_V$ and the structure dependent part
of the axial vector form factor $F_{A,\rm SD} = F_A + Q_\ell f_P / v p_\gamma$,
where $Q_\ell$ is the lepton charge.
Improved simulation techniques,
such as the all mode averaging~\cite{A2A} and $Z_2$ noise wall-source,
enable them to calculate both $F_V$ and $F_{A, \rm SD}$
in the  full kinematical region
as shown in the right panel of Fig.~\ref{fig:Ld:fB*+rad}.
After their continuum and chiral extrapolation are completed, 
it would be interesting to make a detailed comparison with experiments.


\section{FCNC processes}


The $B$ meson processes mediated by the FCNC,
such as the neutral $B_{(s)}$ meson mixing
and $B_s\!\to\!\mu\mu$, 
occur only beyond the tree-level in the SM,
and hence are sensitive to new physics.
The long-standing tension in the angular distribution of $B\!\to\!K^*\ell\ell$
shown in the right panel of Fig.~\ref{fig:SlD:B2rho}
is a famous example of the B anomalies.
While it is not straightforward to simulate 
this non-gold-plated decay on the lattice,
it would be interesting to extend the study of
$B\!\to\!\rho(\to\!\pi\pi)\ell\nu$ discussed above.
There has been a concern that the tension is due to the insufficient
understanding of the non-perturbative effects from the nearby
charmonium resonances $J/\Psi(1S)$ and $\Psi(2S)$,
namely $B\!\to\!J/\Psi(\to\!\ell\ell)K$ and $\Psi(\to\!\ell\ell)K$.
We note that Ref.~\cite{B2Kll:Nf3:JLQCD:Lat19}
reported a lattice study of the factorization approximation
used to study the long-distance charmonium effects.


\begin{figure}[t]
  \centering
  \begin{minipage}[h]{0.497\linewidth}                 
    \includegraphics[angle=0,width=1.0\linewidth,clip]%
                    {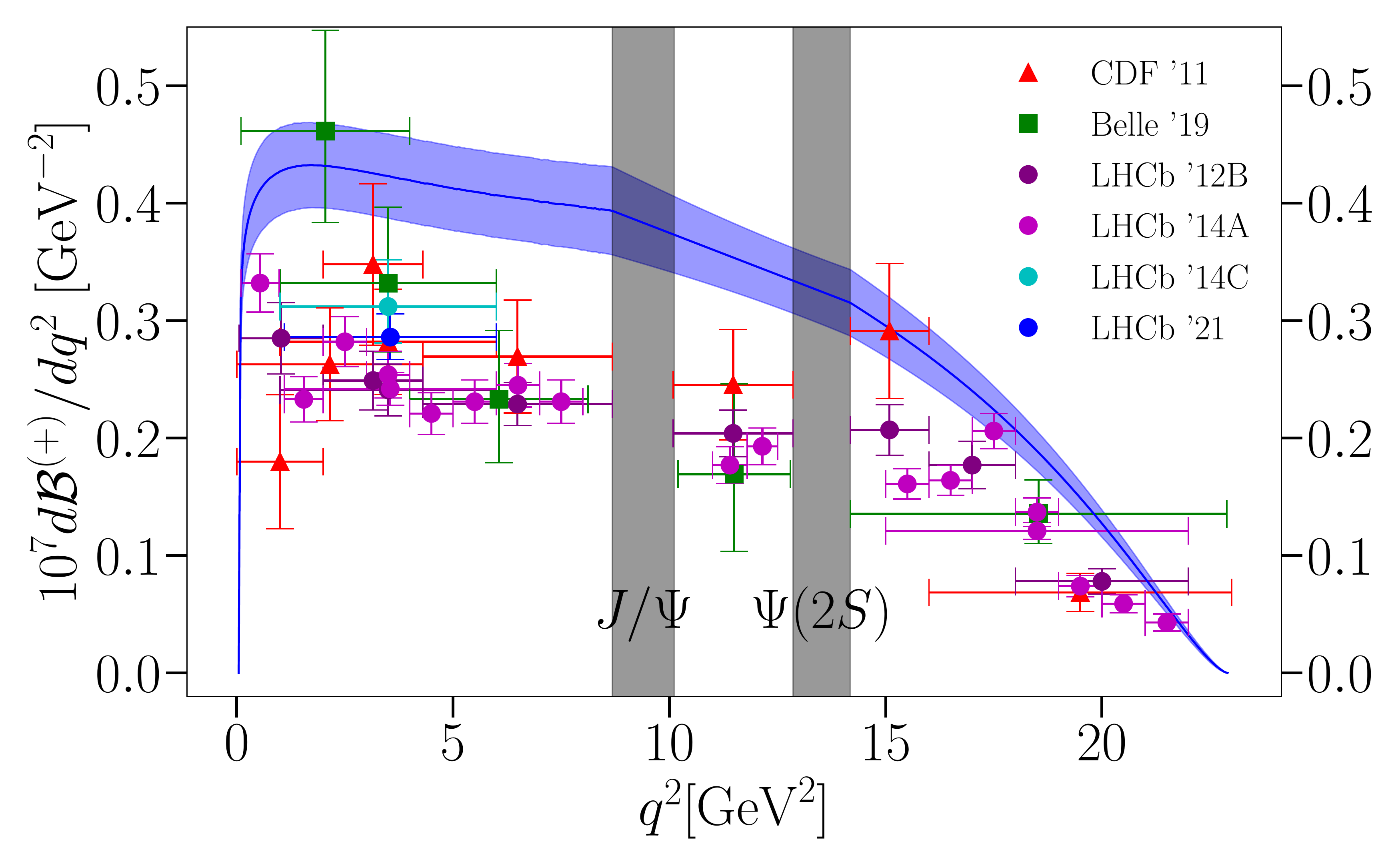}
  \end{minipage}
  \begin{minipage}[h]{0.497\linewidth}                 
    \includegraphics[angle=0,width=1.0\linewidth,clip]%
                    {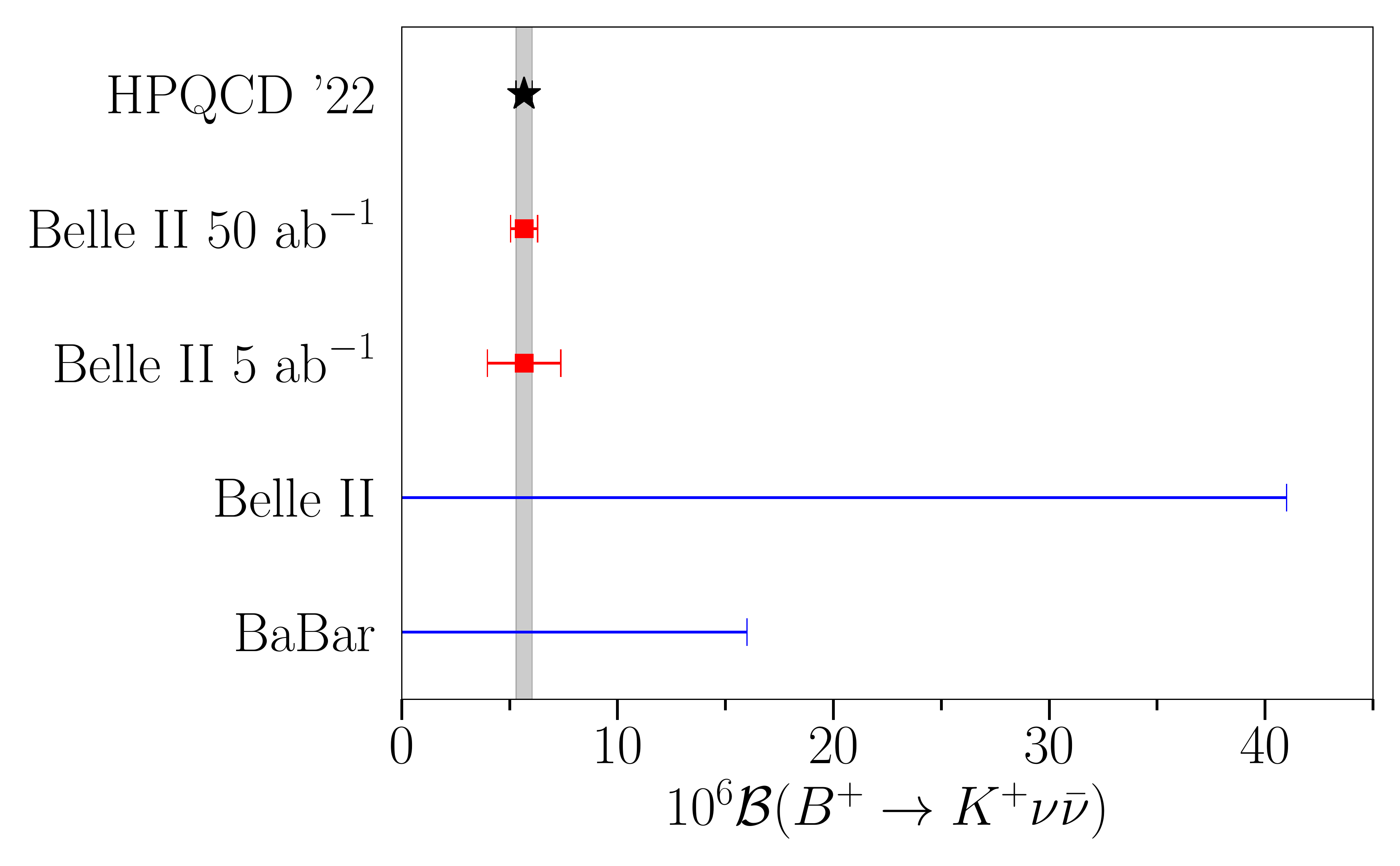}
  \end{minipage}
  \vspace{-2mm}
  \caption{
    Left:
    differential branching fraction $d{\mathcal B}/dq^2$
    for $B\!\to\!K\ell\ell$.
    HPQCD estimate (blue band) is compared with experimental data
    shown by symbols.
    Right:
    branching fraction of $B\!\to\!K\nu\nu$.
    HPQCD estimate is compared with the recent Belle and Belle II measurements
    as well as the expected accuracy for Belle II.
    (Both panels are from Ref.~\cite{B2Kll:Nf4:HPQCD:pheno}.)
  }
  \vspace{-3mm}
  \label{fig:FCNC:B2Kll+nunu}
\end{figure}

On the other hand, it is straightforward to calculate the form factors for
the gold-plated $B\!\to\!K\ell\ell$ decay.
Recently, HPQCD performed a relativistic calculation of 
the vector, scalar and tensor form factors~\cite{B2Kll:Nf4:HPQCD:lat,B2Kll:Nf4:HPQCD:pheno,B2Kll:Nf4:HPQCD:lat22}
\bea
   \langle K(p^\prime) | V_\mu | B(p) \rangle
   & = &
   \left\{ P - \frac{\Delta M^2}{q^2}q \right\}_\mu f_+(q^2)
   + \frac{\Delta M^2}{q^2}q_\mu f_0(q^2),
   \\
   \langle K(p^\prime) | T_{k0} | B(p) \rangle
   & = &
   \frac{2i p_0 p^\prime_k}{M_B+M_K} f_T(q^2),
\eea
where $P\!=\!p+p^\prime$, $q\!=\!p-p^\prime$ and $\Delta M^2\!=\!M_B^2 - M_K^2$.
With a wide range of parameters such as
five lattice cutoffs up to $a^{-1}\!\sim\!4.5$~GeV,
pion masses down to the physical point $\Mpiphys$,
and $m_b$ close to its physical value $\mbphys$
($m_b/m_{b,\rm phys}\!\approx\!0.9$),
they calculate the form factors in the full kinematical range of $q^2$
with the typical accuracy of 4-7\,\%,
where the largest error comes from the statistics.
Fermilab/MILC previously reported that 
theoretical estimate of the differential branching fraction
$d{\mathcal B}(B\!\to\!K\ell\ell)/dq^2$
is systematically larger than experiment~\cite{B2Kll:Nf3:Fermilab/MILC}.
The previous 2\,$\sigma$ tension is enhanced to 4.7\,$\sigma$
with the precise HPQCD results shown in Fig.~\ref{fig:FCNC:B2Kll+nunu}.
This favors a significant shift of the Wilson coefficients for
the effective Hamiltonian operators
$Q_9\!=\!(\bar{q}_L \gamma_\mu b_L)(\bar{\ell}\gamma^\mu \ell)$ and
$Q_{10}\!=\!(\bar{q}_L \gamma_\mu b_L)(\bar{\ell}\gamma^\mu \gamma_5\ell)$
from the SM.


The vector form factor describes the short distance contribution
to the $B\!\to\!K\nu\nu$ decay as
$d{\mathcal B}(B\!\to\!K\nu\nu)/dq^2|_{\rm SD} \!\propto\! |V_{tb}V_{ts}^*|^2 |{\bf p}^\prime|^3 f_+(q^2)^2$.
This decay can be used for the dark sector search,
since it shares missing energy signature with
$B\!\to\!KX_{\rm dark}$ with invisible particle(s) $X_{\rm dark}$.
It is indeed suggested that Belle II is sensitive to dark scalar particles
in the GeV range~\cite{B2Kll:BelleIIprospect}.
The right panel of Fig.~\ref{fig:FCNC:B2Kll+nunu} shows that
i) recent Belle and Belle II measurements of the branching fraction
is consistent with zero,
ii) but it will be largely improved by Belle II,
iii) and the theoretical prediction with the HPQCD's estimate of $f_+$
is already as accurate as the Belle II target accuracy.


\section{$D_{(s)}$ meson decays}


Recent precision lattice calculations of the $K\!\to\!\pi\ell\nu$ form factors
and decay constant ratio $f_K/f_\pi$ as well as better understanding of the
radiative corrections to the superallowed nuclear $\beta$ decays
enable a precision test of CKM unitarity in the first row,
which suggests $\gtrsim\!3\,\sigma$ tension
called the ``Cabibbo angle anomaly''~\cite{FLAG5,Kaon22:Kaneko}.
It is natural to extend the precision test to the second row.
Previously, the accuracy of $|V_{cs(d)}|$ from the $D\!\to\!K(\pi)\ell\nu$ semileptonic decay
is limited by lattice calculation of the relevant form factors.
On the other hand, experimental uncertainty dominates the error of $|V_{cs(d)}|$
from the leptonic decays due to the helicity suppression~\cite{CKM16:WG1}.


\begin{figure}[t]
  \centering
  \begin{minipage}[h]{0.495\linewidth}                 
    \includegraphics[angle=0,width=1.0\linewidth,clip]%
                    {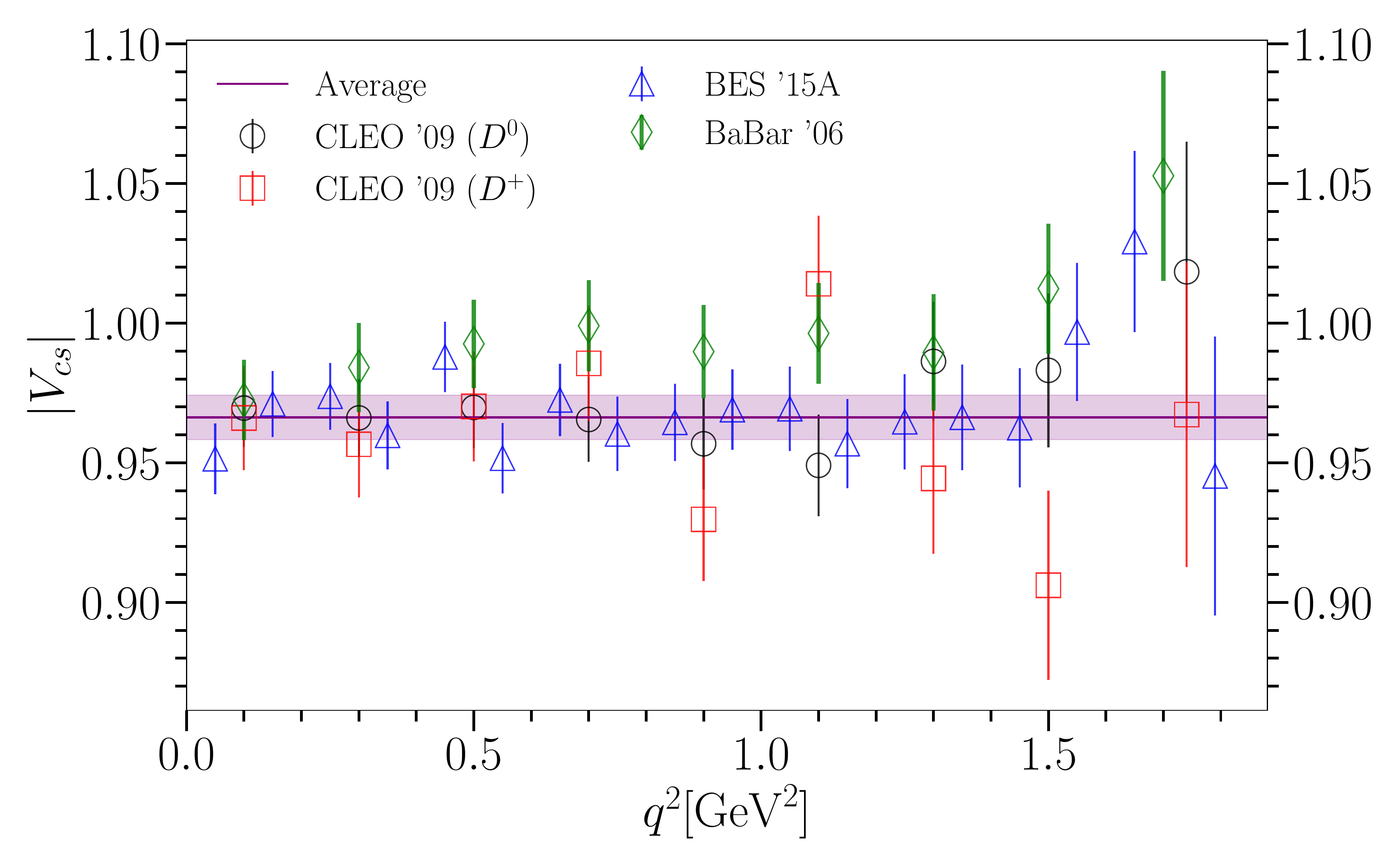}
  \end{minipage}
  \begin{minipage}[h]{0.495\linewidth}                 
    \includegraphics[angle=0,width=1.0\linewidth,clip]%
                    {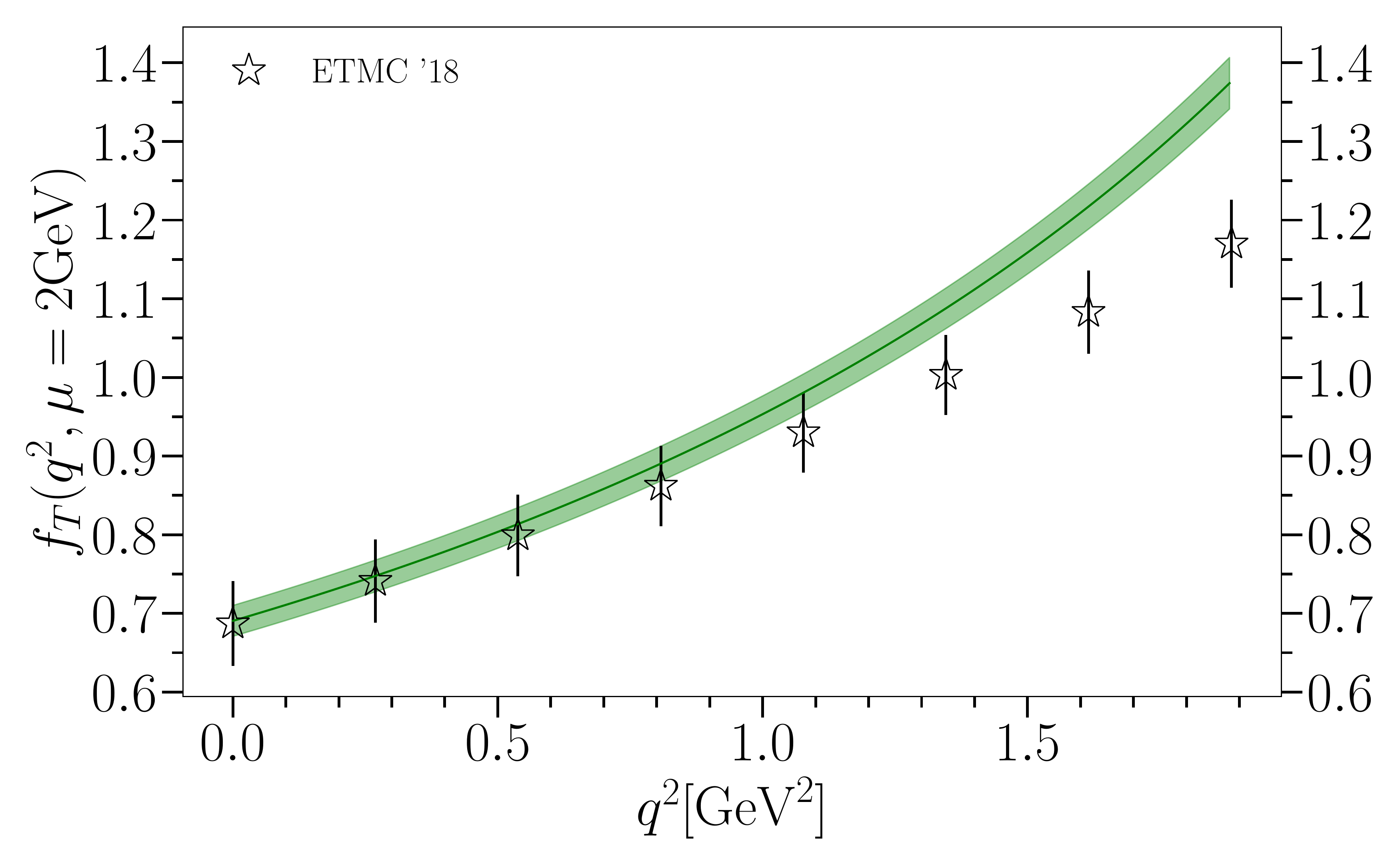}
  \end{minipage}
  \vspace{-2mm}
  \caption{
    Left:
    $|V_{cs}|$ at $q^2$ bins from comparison of differential decay rate
    between theory and experiment
    (figure from Ref.~\cite{D2Kellnu:Nf4:HPQCD}).
    Right:
    tensor form factor $f_T$ as a function of $q^2$
    (figure from Ref.~\cite{B2Kll:Nf4:HPQCD:lat}).
  }
  \vspace{-3mm}
  \label{fig:charm:hpqcd}
\end{figure}

Recently, HPQCD carried out a precision calculation of the $D\!\to\!K\ell\nu$
form factors by a fully realistic simulation:
namely, at five lattice cutoffs up to $\lesssim\!4.5$~GeV
and at physical light, strange and charm quark masses~\cite{B2Kll:Nf4:HPQCD:lat,D2Kellnu:Nf4:HPQCD}.
As suggested by the left panel of Fig.~\ref{fig:charm:hpqcd},
the form factor shape is consistent with experimental data,
and $|V_{cs}|$ is determined with the accuracy of 1\,\%,
where the theoretical and experimental uncertainties are comparable to each other.
As a result, CKM unitarity is confirmed with an accuracy of about 4\,\% as
$|V_{cd}|^2+|V_{cs}|^2 + |V_{cb}|^2\!=\!0.998(35)_{cd}(13)_{cs}$~\cite{FLAG5},
where $|V_{cb}|$ is too small to significantly contribute to this test.


A more precise determination of $|V_{cd}|$ is the next target of the unitarity test.
To this end, it is encouraging that RBC/UKQCD~\cite{D2piellnu:Nf3:RBC/UKQCD},
ALPHA/CLS\cite{D2piellnu:Nf3:ALPHA/CLS} and
Fermilab/MILC\cite{D2piellnu:Nf3:Fermilab/MILC}
collaborations reported their preliminary and/or blinded analyses of
$D\!\to\!\pi\ell\nu$.
We note that Refs.~\cite{D2piellnu:Nf3:RBC/UKQCD,D2piellnu:Nf3:Fermilab/MILC}
also discuss an alternative determination of $|V_{cd}|$
through $D_s\!\to\!K\ell\nu$, which is advantageous
on the lattice as discussed above.



There is, however, a concern about the $D\!\to\!K$ tensor form factor.
As shown in the right panel of Fig.~\ref{fig:charm:hpqcd},
there is about 3\,$\sigma$ disagreement between recent determinations
by HPQCD~\cite{B2Kll:Nf4:HPQCD:lat} and ETM~\cite{D2K:tensor:Nf4:ETM}
around $q^2_{\rm max}$, where the lattice calculations are expected
to be reliable. At this moment, this disagreement is not understood,
and more independent studies are necessary.


\section{Summary}

In this article, we reviewed
recent progress on heavy flavor physics from lattice QCD.
%
%
There has been steady progress in precisely calculating
the gold-plated quantities by realistic simulations.
A good accuracy has been already achieved for
some of the experimentally challenging processes, for instance,
helicity-suppressed leptonic decays (decay constants)
and loop-suppressed FCNC decays ($B\!\to\!K\ell\ell,K\nu\nu$ form factors).
For the exclusive semileptonic decays, on the other hand,
more realistic simulations at physical quark masses
and at large recoils as well as careful investigation of systematics,
such as the ground state saturation, are needed to resolve
the $|V_{ub}|$ and $|V_{cb}|$ tensions and to search for new physics.
These are also helpful in resolving the existing tensions
among lattice studies
about $f_{B^*}$, $B_s\!\to\!K$ form factors at low $q^2$
and $D\!\to\!K$ tensor form factor at large $q^2$
likely due to inadequate understanding of systematics.


We stress that there has been remarkable progress in developing
new applications to non-gold-plated processes.
In particular, the inclusive decays attract much attention
and the feasibility and systematics are being actively studied.
Note also that the currently developed methods have wide applications
beyond $B$ physics, for instance, to
the inclusive $\tau$ decay and $\ell N$ scattering.
However, taking the infinite volume limit has not been tested well,
and is important future subject towards realistic simulation of the
inclusive $B$ decays.
There has been good progress also in the study on
$B\!\to\!\ell\nu\gamma$, $B\!\to\!\rho(\to\!\pi\pi)\ell\nu$
and life time difference.
The remaining challenging subjects include 
an extension to long distance contributions to $B\!\to\!K^*\ell\ell$,
$D$ mixing and QCD based study of the factorization.

\vspace{3mm}

I thank
O.~B\"ar. A.~Barone, A.~Broll, M.~Black, C.T.H.~Davies, J.~Frison, D.~Giusti, 
S.~Hashimoto, W.~Jay, A.~J\"uttner, R.~Kellermann, T.~Kitahara, E.~Kou, J.~Lin,
L.~Leskovec, A.~Lytle, M.~Marshall, S.~Mishima, W.G.~Parrott, A.~Smecca,
R.~Sommer, A.~Soni, A.~Vaquero and L.~Vittorio
for their input and discussions.
This work is supported in part by JSPS KAKENHI Grant Number 21H01085
and by the Toshiko Yuasa France Japan Particle Physics Laboratory
(TYL-FJPPL project FLAV\_03).

\end{document}